\documentclass[12pt,fleqn]{article}
\usepackage{hyperref}
\usepackage{epsfig}
\usepackage{latexsym}
\usepackage{graphicx}
\usepackage{amssymb,amsmath}
\usepackage{amsthm}
\usepackage[x11names]{xcolor}
\usepackage{cite}
\usepackage{subcaption}
\usepackage{caption}
\usepackage{float}
\usepackage{mathtools}
\usepackage{tabularx}
\usepackage{comment}

\newtheorem{theorem}{Theorem}

\newtheorem{cor}[theorem]{Corollary}
\newtheorem{assumption}[theorem]{Assumption}
\newtheorem{proposition}[theorem]{Proposition}
\newtheorem{rem}[theorem]{Remark}
 
\title{Simplified model of immunotherapy for glioblastoma multiforme: cancer stem cells hypothesis perspective}
\author{Wiktor Jochymczyk, \ Urszula Fory\'s}
\date{February 9, 2025}

{
}

\begin{document}
\maketitle
    
\begin{abstract}
Despite ongoing efforts in cancer research, a fully effective treatment for glioblastoma multiforme (GBM) is still unknown. Since adoptive cell transfer immunotherapy is one of the potential cure candidates, efforts have been made to assess its effectiveness using mathematical modeling. In this paper, we consider a model of GBM immunotherapy proposed by Abernathy and Burke (2016), which also takes into account the dynamics of cancer stem cells, i.e., the type of cancer cells that are hypothesized to be largely responsible for cancer recurrence. We modify the initial ODE system by applying simplifying assumptions and analyze the existence and stability of steady states of the obtained simplified model depending on the treatment levels.
\end{abstract}

\section{Introduction}
Glioblastoma multiforme (GBM) accounts for $15.6\%$ of all and up to
$45.2\%$ of malignant primary brain tumors~\cite{GBM1}. Due to its sensitive location, high heterogeneity and strong suppression of the immune response, the therapy that allows for a complete cure is still unknown~\cite{Hetero,Supress}. Among patients diagnosed with GBM, the median survival time from diagnosis is 15 months and less than $5\%$ of patients survive $5$ years from diagnosis~\cite{GBM1, GBM2}.

The commonly accepted treatment for GBM consists of tumor resection followed by radiotherapy combined with temozolomide chemotherapy~\cite{GBM3}. Due to the low effectiveness of conventional therapy, research on alternative treatment methods is still ongoing. These include various types of immunotherapy, such as treatment with cancer vaccines, oncolytic viruses, immune checkpoint inhibitors, and adoptive cell transfer~\cite{Immuno}.

In \cite{Kronik}, Kronik et al.~presented a mathematical model for the treatment of GBM by adoptive transfer of cytotoxic T cells. That work was then continued by Kogan et al.~in~\cite{Kogan}. More precisely, in~\cite{Kogan}, the model proposed in~\cite{Kronik} was first generalized by assuming certain properties of the functions that describe the processes considered in the original article and reflected in that article by specific functions. For this generalized model, Kogan {\it et al.} proved the boundedness of solutions and the dissipativity of the model. Since every dissipative system has a global attractor, steady states were studied as candidates for it. In the general case, the steady states established for the untreated system were considered. Then, for the system with treatment, conditions for tumor elimination were proposed. The final section was devoted to confronting analytical and numerical results with clinical data.

In \cite{Abernathy}, Abernathy and Burke built on previous results and modified the general model from~\cite{Kogan} to include cancer stem cells (CSCs). The CSC hypothesis states that a special type of cancer cells with properties similar to healthy stem cells is largely responsible for the formation and development of cancer. These cells are thought to multiply much faster than normal cancer cells and then differentiate into other types of cancer cells. Theoretically, if even a small number of CSCs survive treatment, it could lead to relapse.

The GBM development and immunotherapy model considered in this article is based on the ODE system presented in \cite{Abernathy}. First, we adopt the notation used in that article:

\begin{itemize}
    \item $T$ and $S$ will refer to the number of ordinary cancer cells and CSCs, respectively;
    \item $C$ will be the number of cytotoxic T lymphocytes that fight the tumor;
    \item $x$, $y$ will refer to the amount of transforming growth factor beta (TGF-$\beta$) and the amount of interferon gamma (IFN-$\gamma$) in the vicinity of the tumor, respectively;
    \item $u$ and $v$ will be the average number of major histocompatibility complex molecules of class I and II (MHC class I and II) per tumor cell, respectively.
\end{itemize}

With the above notation, the model proposed in \cite{Abernathy} has the following form:

\begin{subequations} \label{eq:system}
\renewcommand{\theequation}{\theparentequation.\arabic{equation}}
  \begin{align}
      \dot{T} &= \alpha(T, S) + R_T(T)T - f_{T}(x)g_{T}(u)h_{T}(T)CT,\label{eq:system_1}\\ 
      \dot{S} &= R_S(S)S - \alpha(T, S) - f_{S}(x)g_{S}(u)h_{S}(S)CS,\label{eq:system_2}\\
      \dot{C} &= f_{C}((T + S)v)g_{C}(x) - \mu_{C}C + N(t),\label{eq:system_3}\\ 
      \dot{x} &= f_{x}(T, S) - \mu_{x}x,\label{eq:system_4}\\
      \dot{y} &= f_{y}(C) - \mu_{y}y,\label{eq:system_5}\\
      \dot{u} &= f_{u}(y) - \mu_{u}u,\label{eq:system_6}\\
      \dot{v} &= f_{v}(x)g_{v}(y) - \mu_{v}v.\label{eq:system_7}
  \end{align}
\end{subequations}

We will now present interpretations of the equations of System~\eqref{eq:system}. Equation~\eqref{eq:system_1} describes the change in the number of tumor cells over time. The function $\alpha$ is responsible for the process of stem cells differentiating into regular cancer cells, while the function $R_T$ describes the growth of the tumor cell population \textit{per capita}. The third term reflects the efficiency with which lymphocytes destroy cancer cells. The functions $f_{T}$, $g_{T}$, and $h_{T}$ introduce a dependence of this process on the amount of TGF-$\beta$ in the vicinity of the tumor, the average number of MHC class I molecules per tumor cell and the size of the population of tumor cells, respectively. 

Similarly, Equation~\eqref{eq:system_2} describes the change in the CSC population over time. In this case, the function $R_S$ describes the growth of the CSC population \textit{per capita}, and the last term reflects the efficiency with which lymphocytes destroy CSCs, analogously to Eq.~\eqref{eq:system_1}. The change in the population of lymphocytes is in turn described by Eq.~\eqref{eq:system_3}. The functions $f_{C}$, $g_{C}$ make this process dependent on the total number of MHC class II molecules on the surface of all tumor cells and the amount of TGF-$\beta$ in the vicinity of the tumor, respectively. In addition, $\mu_{C}$ is a constant death rate for lymphocytes. The function $N$ reflects the influx of lymphocytes into the vicinity of the tumor at time $t$ as a result of immunotherapy. 

The change in the amount of TGF-$\beta$ over time is expressed by Eq.~\eqref{eq:system_4} where the function $f_{x}$ describes the production of TGF-$\beta$ by both types of cancer cells, and the constant $\mu_{x}$ reflects the intensity with which TGF-$\beta$ particles decay over time. Equation~\eqref{eq:system_5} describing the change in the amount of IFN-$\gamma$ is analogous, but this time lymphocytes $C$ are responsible for the production. Similarly, Equation~\eqref{eq:system_6} describes the change in the average number of MHC class I molecules on the surface of cancer cells, where the function $f_{u}$ reflects the influence of IFN-$\gamma$ on this process, and $\mu_{u}$ is the coefficient with which MHC class I molecules disappear over time. Finally, Equation~\eqref{eq:system_7} describes the change in the number of MHC class II molecules on the surface of cancer cells, where the influence of TGF-$\beta$ and IFN-$\gamma$ on this process is reflected in functions $f_{v}$ and $g_{v}$, respectively.

The model presented above is the subject of analysis performed in the following sections under additional assumptions, which we will pose below.

\section{Basic properties of the model}

\begin{figure}[h]
    \centering
    
    \begin{subfigure}{0.49\textwidth}
    \centering
    \captionsetup{justification=centering,margin=0.5cm}    
    \includegraphics[width=1\textwidth]{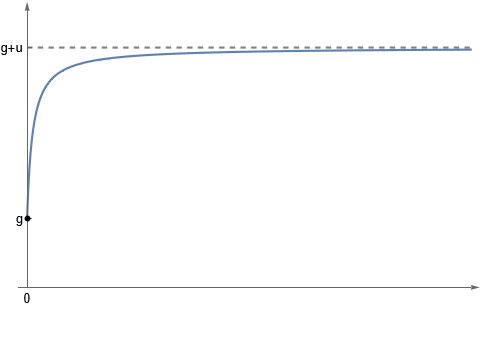}
    \caption{$f_u,$ $f_C.$}
    \label{fig:0-a}
    \end{subfigure}
    \hfill
    \begin{subfigure}{0.49\textwidth}
    \centering
    \captionsetup{justification=centering,margin=0.5cm}
\includegraphics[width=1\textwidth]{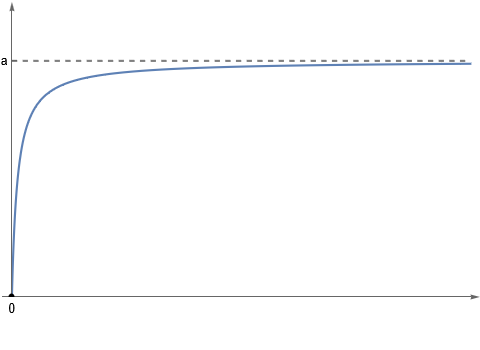}
    \caption{$g_T,$ $g_S,$ $f_C,$ $g_v.$}
    \label{fig:0-b}
    \end{subfigure}
    
    \begin{subfigure}{0.49\textwidth}
    \centering
    \captionsetup{justification=centering,margin=0.5cm}    
    \includegraphics[width=1\textwidth]{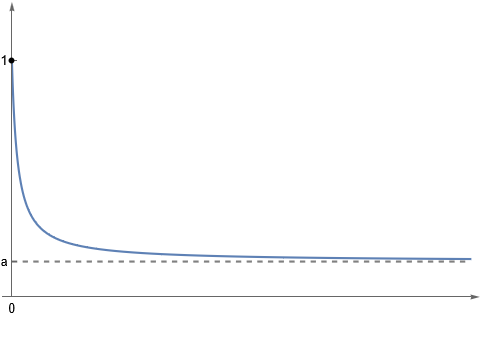}
    \caption{$f_T,$ $f_S,$ $g_C.$}
    \label{fig:0-c}
    \end{subfigure}
    \hfill
    \begin{subfigure}{0.49\textwidth}
    \centering
    \captionsetup{justification=centering,margin=0.5cm}
\includegraphics[width=1\textwidth]{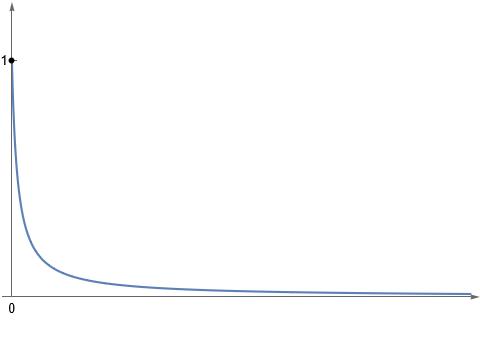}
    \caption{$h_T,$ $h_S,$ $f_v.$}
    \label{fig:0-d}
    \end{subfigure}
    
    \caption{Exemplary graphs of some of the functions described in Assumption \ref{assumption1}.}
    \label{fig:0}
\end{figure}

In this section, we discuss the basic properties of the model. To this end, we need to specify assumptions about the functions describing its RHS.
\begin{assumption}\label{assumption1}
From now on we assume that:

\begin{enumerate}
    \item all the functions $\alpha$, $R_{i}$, $f_{i}$, $g_{i}$, $h_{i}$, $i=T$, $S$, $f_{C}$, $g_{C}$, $f_{x}$, $f_{y}$, $f_{u}$, $f_{v}$ and $g_{v}$ are of class $\mathbf{C}^{1}$ and are nonnegative for biologically acceptable ranges of the variables;
    \item $T_{0}$, $S_{0}$, $C_{0}$, $x_{0}$, $y_{0}$, $u_{0}$, $v_{0}\geqslant 0$; moreover, $T_0\leqslant K_T$, $S_0\leqslant K_S$, where $K_T$, $K_S$ are the maximal sizes of cellular populations;
    \item $R_i(Z)=r_i \left(1-\frac{Z}{K_i}\right)$, $i=T$, $S$, where $r_T$, $r_S>0$ are maximal proliferation rates for  the populations;
    \item $\alpha(T,S) = r_{\alpha}\frac{S}{K_S}\frac{T}{K_T}(K_T-T)$, where $r_{\alpha}>0$ is some constant;
    \item $f_{i}$, $i=T$, $S$, are decreasing, $f_{i}(0) = 1$ and $\lim\limits_{x\to\infty} f_{i}(x)=a_{i,x}>0$,  where $a_{T,x}$, $a_{S,x}$ are some constants (see Fig.~\ref{fig:0-c});
    \item $g_{i}$,  $i=T$, $S$, are increasing, $g_{i}(0)  = 0$, $g_{S}(u)<g_{T}(u)$ for $u>0$; \\moreover, $\lim\limits_{u\to\infty} g_{i}(u)$ $= a_{i,u} > 0$, where $a_{T,u}$, $a_{S,u}$ are some constants (see Fig.~\ref{fig:0-b});
    \item $h_{i}$,  $i=T$, ${S}$, are decreasing, $h_{i}(0)  = 1$, $\lim\limits_{Z\to\infty} h_{i}(Z) = 0$ (see Fig.~\ref{fig:0-d});
    \item $f_{C}$ is increasing, $\lim_{z\to\infty}f_{C}(z)=a_{C,v}>0$ for some constant $a_{C,v}$, $f_{C}'(0)>0$ and $\lim\limits_{z\to\infty}f_{C}'(z)=0$ (see Fig.~\ref{fig:0-a},~\ref{fig:0-b});
    \item $g_{C}$ is decreasing, $g_{C}(0)=1$ and $\lim\limits_{x\to\infty}g_{C}(x)=a_{C,x}>0$ for some constant $a_{C,x}$ (see Fig.~\ref{fig:0-c});
    \item $N(t) \equiv N \geqslant0$, i.e.~the therapy is constant in time;
    \item $f_{x}(T,S)=g_{x} + a_{x,T}T + a_{x,S}S$, $f_{y}(C)=a_{y,C}C$, where $g_x, a_{x,T}$, $a_{x,S}$, $a_{y,C} > 0$ are some constants;
    \item $f_u$ is increasing, $f_{u}(0)=g_{u}>0$ and $\lim\limits_{y\to\infty} f_{u}(y) = g_{u} + a_{u,y}$, where $a_{u,y} > 0$ is some constant (see Fig.~\ref{fig:0-a});
    \item $f_{v}$ is decreasing to zero and $f_{v}(0)=1$ (see Fig.~\ref{fig:0-d});
    \item $g_{v}$ is increasing, $g_{v}(0)=0$, $\lim\limits_{y\to\infty} g_{v}'(y)= 0$, $\lim\limits_{y\to\infty} g_{v}(y)=a_{v,y}>0$ for some constant $a_{v,y}$ (see Fig.~\ref{fig:0-b}).
\end{enumerate}
\end{assumption}

From the above assumptions, we find that the RHS of System~\eqref{eq:system} is of class $\mathbf{C}^{1}$, which guarantees the local existence and uniqueness of solutions. In \cite{Abernathy}, following the ideas of Kogan {\it et al.},~\cite{Kogan},  the authors proved the dissipativity of the System \eqref{eq:system} and then analyzed the tumor-free steady state. However, they obtained implicit conditions of the stability, which we would like to reformulate in explicit form. For further analysis of tumor persistence states, they posed a simplifying assumption which we also exploit in our paper.

Because System~\eqref{eq:system} is highly complex, for its further analysis we introduce the following simplifying assumption.

\begin{assumption}\label{assumption2}
We assume that $N$ is the total number of lymphocytes that enter the tumor area as a result of immunotherapy and the body's natural immune response, i.e.~Eq.~\eqref{eq:system_3} reads
$$\dot{C}=N-\mu_{C}C.$$
\end{assumption}

The assumption above implies that $v$ no longer influences other variables in the system. Thus, we can consider System~\eqref{eq:system} without the last equation.
Note that for any initial conditions the variables $C$, $y$, and $u$ converge exponentially to $C^*(N),$ $y^*(N),$ $u^*(N)$, where

\begin{equation*}
    C^*(N):=\frac{N}{\mu_{C}}, \; y^*(N):= \frac{f_{y}\big(C^*(N)\big)}{\mu_{y}}, \; u^*(N):= \frac{f_{u}\big(y^*(N)\big)}{\mu_{u}}.
\end{equation*}

Thus, we will consider a quasi-stationary approximation for $C,y,u$. Eventually, System~\eqref{eq:system} takes the following form:

\begin{subequations}\label{eq:our_system}
\renewcommand{\theequation}{\theparentequation.\arabic{equation}}
    \begin{align}
        \dot{T} &= r_{\alpha}\frac{S}{K_S}\frac{T}{K_T}(K_T-T) + r_T \left(1-\frac{T}{K_T}\right)T - f_{T}(x)G_T(N)h_{T}(T)T, \label{eq:our_system_1}\\
        \dot{S} &= r_S \left(1-\frac{S}{K_S}\right)S - r_{\alpha}\frac{S}{K_S}\frac{T}{K_T}(K_T-T) - f_{S}(x)G_S(N)h_{S}(S)S,\label{eq:our_system_2}\\
        \dot{x} &= g_{x} + a_{x,T}T + a_{x,S}S - \mu_{x}x,\label{eq:our_system_3}
    \end{align}
\end{subequations}
where
\begin{equation*}
    G_i(N):=g_{i}\big(u^*(N)\big)C^*(N),
    \quad i=T, \, S.
\end{equation*}

\begin{rem}
Under Assumption~\ref{assumption2} the stability of steady states of the full System~\eqref{eq:system} and the simplified System~\eqref{eq:our_system} is equivalent.
\end{rem}

The above remark allows us to note that without loss of generality it is enough to study steady states of the simplified system instead of the full one.

Now we proceed to the analysis of the properties of the functions $G_i(N)$, $i=T$, $S$. Based on Assumption \ref{assumption1} we get that
\begin{equation*}
    \begin{aligned}
    u^*(N)=&
    \frac{f_{u}(y^*)}{\mu_{u}}=
    \frac{1}{\mu_u}f_{u}\left(\frac{a_{y,C}C^*(N)}{\mu_y}\right)=
    \frac{1}{\mu_u}f_{u}\left(\frac{a_{y,C}N}{\mu_y\mu_C}\right)
    \end{aligned}
\end{equation*}
and thus
\begin{equation*}
    G_i(N)= g_{i}\left(\frac{1}{\mu_u}f_{u}\left(\frac{a_{y,C}N}{\mu_y\mu_C}\right)\right)\frac{N}{\mu_{C}},
\quad i=T,\,S.
\end{equation*}
Since the functions $f_u$ and $g_i$, $i=T$, $S$, are increasing and nonnegative, so are the functions $G_i$, $i=T$, $S$. Moreover,
\begin{equation*}
    G_i(N)=0 \iff N=0, \quad i=T,\, S,
\end{equation*}
and
\begin{equation*}
    \lim_{N\rightarrow\infty}G_i(N)=\infty, \quad i=T,\, S.
\end{equation*}
Because $\lim_{u\rightarrow\infty}g_i(u)=a_{i,u}$ and 
$g_i$ are increasing, $i=T$, $S$,
we get that
\begin{equation*}
     \lim_{N\rightarrow\infty}\frac{G_i(N)}{N}=
\lim_{N\rightarrow\infty}g_{i}\left(\frac{1}{\mu_u}f_{u}\left(\frac{a_{y,C}N}{\mu_y\mu_C}\right)\right)\frac{1}{\mu_{C}}\leqslant \frac{a_{i,u}}{\mu_C},\quad  i=T,\, S.
\end{equation*}
Furthermore, since $\forall{u>0}\;\;g_T(u)>g_S(u),$ then the following corollary holds.

\begin{cor}\label{cor:G1_G2}
If Assumptions \ref{assumption1} and \ref{assumption2} are satisfied, then the functions $G_i$,  $i=T$, $S$, are strictly increasing and linearly bounded. Moreover
\begin{equation*}
G_T(0)=G_S(0)=0 \quad \text{and} \quad
    \forall{N>0}
\ \ G_T(N)> G_{S}(N) .
\end{equation*}
\end{cor}

\begin{proposition}
If Assumption~\ref{assumption1} is satisfied, then System~\eqref{eq:our_system} has unique nonnegative solutions for nonnegative initial data. Moreover, if $T_0\leq K_T$ and $S_0\leq K_S$, then the solutions are bounded from above and defined for all $t\geqslant 0.$
\end{proposition}

\textit{Proof:} Existence and uniqueness are a simple consequence of Assumption~\ref{assumption1}. We will now show that the solutions are nonnegative. From Equation~\eqref{eq:our_system_1} we get that if $T=0$, then $\dot{T}=0$. Hence, if $T_0=0$, then $T(t)=0$ for all $t\geqslant 0$ for which the solution exists. The uniqueness of the solutions implies that if $T_0>0$, then $T(t)>0$ for all $t\geqslant 0$ for which the solution exists. The same argument applies for Eq.~\eqref{eq:our_system_2}.

Now, consider Eq.~\eqref{eq:our_system_3}. Because for $x=0$ there is $\dot x = g_{x} + a_{x,T}T + a_{x,S}S \geq g_x>0 $, the variable $x$ cannot take negative values for $x_0\geq 0$.
Thus, all the coordinates of the  solutions to System~\eqref{eq:our_system} are nonnegative.

Next, we proceed to show that the solutions to System \eqref{eq:our_system} are also bounded from above. Note that
\begin{equation}\label{eq:T_bound}
    \begin{aligned}
        \dot{T} &\leqslant r_{\alpha}\frac{S}{K_S}\frac{T}{K_T}(K_T-T) + r_T \left(1-\frac{T}{K_T}\right)T \\
        &= \left( r_{\alpha}\frac{S}{K_S} + r_T\right)\frac{T(K_T-T)}{K_T}.
    \end{aligned}
\end{equation}
Let us assume that $T_0\in [0,K_T]$ and there exists $t_1$ such that $T(t_1)=K_T$ and $T$ exceeds $K_T$ at $t_1$. Then there must exist such $\epsilon > 0$ that for each $t\in(t_1,t_1+\epsilon)$ $T(t)>K_T.$ From \eqref{eq:T_bound} we, therefore, get $\dot{T}(t)<0$ for all $t\in(t_1,t_1+\epsilon)$. On the other hand, from the Lagrange theorem there exists such $\xi \in (t_1,t_1+\epsilon)$ that
\begin{align*}
    \dot{T}(\xi)=&\frac{T(t_1+\epsilon)-T(t_1)}{\epsilon}=\frac{T(t_1+\epsilon)-K_T}{\epsilon}>0,
\end{align*}
which yields a contradiction. Thus, $T\leqslant K_T$, and hence we have
\begin{equation*}
    \begin{aligned}
        \dot{S} &= r_S \left(1-\frac{S}{K_S}\right)S - r_{\alpha}\frac{S}{K_S}\frac{T}{K_T}(K_T-T) - f_{S}(x)G_{S}(N)h_{S}(S)S \\
        &\leqslant  r_S \left(1-\frac{S}{K_S}\right)S.
    \end{aligned}
\end{equation*}
This means that $S$ does not exceed solutions of the logistic equation $$\dot{S}=r_S\left(1-\frac{S}{K_S} \right)S,$$ which implies $S\leqslant K_S$. Thus, we have
\begin{align*}
    \dot{x}=&g_{x} + a_{x,T}T + a_{x,S}S-\mu_x x\leqslant g_x + a_{x,T}K_T + a_{x,S}K_S - \mu_x x.
\end{align*}
Let us define
\begin{align*}
    \bar{x}:=\frac{g_x+a_{x,T}K_T+a_{x,S}K_S}{\mu_x}.
\end{align*}
It is easy to see that if for a given $t\geqslant 0$ we have $x(t)>\bar{x}$, then $\dot{x}(t)<0$. Hence, if $x_0>\bar{x}$, then $x(t)\leqslant x_0$ for all $t$ for which the solution exists. If $x_0\leqslant\bar{x},$ then $x(t)\leqslant \bar{x}$. Thus, $x$ is bounded from above by $x_{\max}:=\max\{x_0, \bar{x}\}$.

Finally, it is easy to show that since $T,$ $S$ and $x$ are bounded from above, then their derivatives are also bounded. Hence, the solutions to System \eqref{eq:our_system} exist for all $t\geqslant 0$.
$\hfill\square$

\begin{cor}
If Assumption \ref{assumption1} is satisfied, then the System~\eqref{eq:our_system} has an invariant set defined by $\mathcal{D}=[0,K_T]\times[0,K_S]\times[0,\bar x]$. 
\end{cor}

\section{Analysis of steady states}

In this section we will analyze the existence and uniqueness of selected steady states of System~\eqref{eq:our_system}, depending on the value of $N$. The steady states of System~\eqref{eq:our_system} solve the following system of algebraic equations
\begin{subequations}\label{eq:our_system_steady}
\renewcommand{\theequation}{\theparentequation.\arabic{equation}}
    \begin{align}
        0 &= r_{\alpha}\frac{S^*}{K_S}\frac{T^*}{K_T}(K_T-T^*) + r_T \left(1-\frac{T^*}{K_T}\right)T^* - f_{T}(x^*)G_{T}(N)h_{T}(T^*)T^*, \label{eq:our_system_steady_1}\\
        0 &= r_S \left(1-\frac{S^*}{K_S}\right)S^* - r_{\alpha}\frac{S^*}{K_S}\frac{T^*}{K_T}(K_T-T^*) - f_{S}(x^*)G_{S}(N)h_{S}(S^*)S^*,\label{eq:our_system_steady_2}\\
        0 &= g_{x} + a_{x,T}T^* + a_{x,S}S^* - \mu_{x}x^*.\label{eq:our_system_steady_3}
    \end{align}
\end{subequations}
In particular, if a point $(T^*,S^*,x^*)$ is a steady state of System~\eqref{eq:our_system}, then 
\begin{equation}\label{eq:x_steady}
 x^* = \frac{g_x + a_{x,T}T^* + a_{x,S}S^*}{\mu_x}.
\end{equation}
To analyze the stability of the steady states we will use the linearization theorem. Let us denote the RHS of Eqs.~\eqref{eq:our_system_1} and \eqref{eq:our_system_2} as $f_1$ and $f_2$, respectively. Then the Jacobian matrix $\mathcal{J}$ of System~\eqref{eq:our_system} at any point $(T,S,x)$ reads
\begin{align}\label{jacobi}
    \mathcal{J}(T,S,x)=&
    \begin{pmatrix}
        \frac{\partial f_1}{\partial T}(T,S,x)
        &
        \frac{\partial f_1}{\partial S}(T,S,x)
        &
        \frac{\partial f_1}{\partial x}(T,S,x)
        \\[10pt]
        \frac{\partial f_2}{\partial T}(T,S,x)
        &
        \frac{\partial f_2}{\partial S}(T,S,x)
        &
        \frac{\partial f_2}{\partial x}(T,S,x)
        \\[10pt]
        a_{x,T} & a_{x,S} & -\mu_x
    \end{pmatrix},
\end{align}
where
\begin{subequations}\label{eq:partial_derivatives}
\renewcommand{\theequation}{\theparentequation.\arabic{equation}}
  \begin{align}
    \frac{\partial f_1}{\partial T}(T,S,x)=& 
        \begin{aligned}[t] 
        & \frac{r_\alpha S}{K_T K_S}(K_T - 2T) - \frac{r_T}{K_T}T + r_T \left(1-\frac{T}{K_T}\right) \\ 
        &-f_T(x)G_T(N)\big(h_T'(T)T^* + h_T(T)\big),
        \end{aligned} \label{eq:partial_derivatives_1}
    \\
    \frac{\partial f_1}{\partial S}(T,S,x)=& \, \frac{r_\alpha T}{K_T K_S}(K_T - T),\label{eq:partial_derivatives_2}
    \\
    \frac{\partial f_1}{\partial x}(T,S,x)=& -f_T'(x)G_T(N)h_T(T)T,\label{eq:partial_derivatives_3}
    \\
    \frac{\partial f_2}{\partial T}(T,S,x)=& \,\frac{r_\alpha S}{K_T K_S}(K_T - 2T),\label{eq:partial_derivatives_4}
    \\
    \frac{\partial f_2}{\partial S}(T,S,x)=& 
        \begin{aligned}[t]
        & -\frac{r_S}{K_S}S + r_S \left(1-\frac{S}{K_S}\right) - \frac{r_\alpha T}{K_T K_S}(K_T - T)\\
        &- f_S(x)G_S(N)\big(h_S'(S)S + h_S(S)\big),
        \end{aligned}\label{eq:partial_derivatives_5}
    \\
    \frac{\partial f_2}{\partial x}(T,S,x)=&  -f_S'(x)G_S(N)h_S(S)S.\label{eq:partial_derivatives_6}
  \end{align}
\end{subequations}

In subsequent sections, if the context makes it clear at which point we consider a partial derivative, we will often omit the point from the notation, e.g.~we will write $\frac{\partial f_i}{\partial T}$ instead of $\frac{\partial f_i}{\partial T}(T^*,S^*,x^*)$, $i=1$, $2$.

\subsection{The case without treatment}
We begin our analysis of steady states with the case of $N\equiv 0$. 
In this case System~\eqref{eq:our_system_steady} simplifies to
\begin{subequations}\label{N=0}
\renewcommand{\theequation}{\theparentequation.\arabic{equation}}
    \begin{align}
        0 &= \left(  r_{\alpha}\frac{S^*}{K_S} + r_T 
\right)
\left(1-\frac{T^*}{K_T}\right)T^* ,\label{N=0_1} \\
        0 &= \left( r_S (K_S-{S^*}) - r_{\alpha}\frac{T^*}{K_T}(K_T-T^*)
\right) \frac{S^*}{K_S} , \label{N=0_2}\\
        0 &= g_{x} + a_{x,T}T^* + a_{x,S}S^* - \mu_{x}x^*,
    \end{align}
\end{subequations}
and it is obvious that the last coordinate satisfies Eq.~\eqref{eq:x_steady}, while Eqs.~\eqref{N=0_1} and \eqref{N=0_2} do not depend on $x^*$. Hence, we can analyze only the system for $T^*$ and $S^*$.

We easily see that in a steady state
\begin{itemize}
\item either $T^*=0$ or $T^*=K_T$;
\item for both the values of $T^*$, either $S^*=0$ or  $S^*=K_S$.
\end{itemize}
Hence, there are four steady states:
\[
A_1 =(0,0), \quad A_2 =(0,K_S), \quad A_3=(K_T,0), \quad A_4=(K_T,K_S).
\]
To study stability of these states we use the second minor of $\mathcal{J}$ that reads
\[
J_2(T,S) = 
     \begin{pmatrix}
         \frac{r_\alpha S(K_T - 2T)}{K_T K_S} +
         r_T\left(1-\frac{2T}{K_T}\right)
         &
         \frac{r_\alpha T(K_T - T)}{K_T K_S}
         \\[20pt]
         \frac{r_\alpha S(K_T - 2T)}{K_T K_S}
         &
    r_S\left(1-\frac{2S}{K_S}\right)
         - \frac{r_\alpha T(K_T - T)}{K_T K_S} 
     \end{pmatrix},
\]
and it is easy to see that for every $A_i$, $i=1,2,3,4$, eigenvalues are real and equal to the terms on the diagonal. Therefore, in the coordinates $(T,S)$,
\begin{itemize}
    \item $A_1$ is an unstable node, as  $\lambda_1=r_T>0$, $\lambda_2=r_S>0$;
    \item $A_2$ is a saddle, as $\lambda_1=r_{\alpha}+r_T>0$, $\lambda_2=-r_S<0$;
    \item $A_3$ is a saddle, as $\lambda_1=-r_T<0$, $\lambda_2=r_S>0$;
    \item $A_4$ is a stable node, as $\lambda_1=-r_{\alpha}-r_T<0$, $\lambda_2=-r_S<0$.
\end{itemize}
Moreover, it is also easy to see that the RHS of Eq.~\eqref{eq:our_system_1} for $N=0$ is always positive for $T\in(0,K_T)$, which means that $T(t) \to K_T$ for $t\to \infty$ for any initial $T_0 \in (0,K_T)$. Then, it follows that $S(t)\to K_S$ for $S_0\in (0,K_S)$. Hence, $A_4$ is globally stable, which implies that both populations of tumor cells grow to their maximum sizes.

In the next subsection, we will formulate the conditions that allow for a~cure.

\subsection{Semitrivial cure steady state}

First, we consider the simplest case where $T^*=0$, $S^*=0$, $x^*=\frac{g_x}{\mu_x}$. This complete cure steady state (CCS) exists regardless of the value of $N$. 
From Eqs.~\eqref{jacobi} and~\eqref{eq:partial_derivatives} we have
\begin{align*}
    \mathcal{J}\left(0,0,\frac{g_x}{\mu_x}\right) &= 
    \begin{pmatrix}
    r_T - f_T\left(\frac{g_x}{\mu_x}\right)G_T(N) & 0 & 0\\
    0 & r_S - f_S\left(\frac{g_x}{\mu_x}\right)G_S(N) & 0\\
    a_{x,T} & a_{x,S} & -\mu_{x}\\
    \end{pmatrix}.
\end{align*}
Because $\mathcal{J}\left(0,0,\frac{g_x}{\mu_x}\right)$ is a lower triangular matrix its  eigenvalues are
\begin{align*}
    \lambda_{1}&=r_T - f_T\left(\frac{g_x}{\mu_x}\right)G_T(N), &
    \lambda_{2} &= r_S - f_S\left(\frac{g_x}{\mu_x}\right)G_S(N), &
    \lambda_{3} &= - \mu_x.
\end{align*}
Of course $\lambda_3$ is a negative constant, while $\lambda_1$, $\lambda_2 <0$ if and only if
\begin{equation}\label{eq:G1_G2_triv}
    \frac{r_i}{f_i\left(\frac{g_x}{\mu_x}\right)}<G_i(N), \quad i=T,\, S.
\end{equation}
Since the functions $G_i$, $i=T$, $S$, are strictly increasing, their inverse functions $G_i^{-1}$, exist and are also strictly increasing. Thus, we can define
\begin{equation*}
N_{\min}^{(i)}:=G_i^{-1}\left(\frac{r_i}{f_i\left(\frac{g_x}{\mu_x}\right)}\right), 
\quad i=T,\, S.
\end{equation*}
It is easy to see that if $N>N^{loc}_{\min}:=\max\left\{N_{\min}^{(T)}, N_{\min}^{(S)}\right\}$, then Condition \eqref{eq:G1_G2_triv} is satisfied, and therefore the CCS  is locally asymptotically stable.

As local stability is difficult to interpret in practice, we are more interested in stronger condition of global stability. Considering System~\eqref{eq:our_system} in the invariant set $\mathcal{D}$ we can pose such conditions.

\begin{theorem}
    If $N> N_{\min}^{glob}$, where
    \[ N_{\min}^{glob} = \max \left\{ G_T^{-1} \left(\frac{r_T}{f_T(\bar x)h_T(K_T)}\right), G_S^{-1} \left(\frac{r_S}{f_S(\bar x)h_S(K_S)}\right) \right\}, \], then the CCS is globally stable in $\mathcal{D}$. 
\end{theorem}
\textit{Proof:}
First note that in the invariant set, due to monotonicity of the functions $f_i$, $h_i$, $i=T$, $S$, we have
\[
f_i(x)\geq f_i(\bar x), \quad h_i(i) \geq h_i(K_i),\quad i=T, \, S,
\]
and therefore
\[
\dot S \leq r_S\left(1-\frac{S}{K_S}\right)S - f_S(\bar x)G_S(N)h_S(K_S)S
\leq S\left(r_S- f_S(\bar x)G_S(N)h_S(K_S)\right).
\]
We easily see that if the coefficient $\hat r_S=r_S- f_S(\bar x)G_S(N)h_S(K_S)$ 
 is negative, that is,
$G_S(N)>\frac{r_S}{f_S(\bar x)h_S(K_S)}$,
then $S$ tends to 0 exponentially over time.

Similarly, if $G_T(N)>\frac{r_T}{f_T(\bar x)h_T(K_T)}$, then the coefficient 
$$\hat r_T=r_T- f_T(\bar x)G_T(N)h_T(K_T)<0,$$
and according to the assumption of this theorem, there exists $\bar t>0$ such that
\[
r_{\alpha}\frac{S(t)}{K_S} < \frac{\left|\hat r_T \right|}{2}, \quad \text{for all} \ \  t > \bar t.
\]
Therefore, for $t>\bar t$,
\[
\dot T \leq r_{\alpha}\frac{S}{K_S}{T} +r_TT - f_T(\bar x)G_T(N)h_T(K_T)T <
-\frac{\left|\hat r_T \right|}{2} T ,
\]
implies that $T(t) \to 0$ as well. Then it is obvious that $x(t) \to x^*$.
$\hfill\square$


\vspace*{0.3cm}
\subsection{Recurrence steady state}

In this section, we consider a situation in which ordinary cancer cells $T$ have been destroyed as a result of therapy but some population of cancer stem cells $S$ has survived. In other words, we will analyze the steady states $(0,S^*,x^*)$ such that $S^*>0.$ We will call this state the RSS.

Since we assume that $T^*=0$, then
$x^*=\frac{g_x+a_{x,S}S^*}{\mu_x}$, and 
Eq.~\eqref{eq:our_system_steady_2} for $S^*\ne 0$ reads
\begin{equation}\label{eq:steady_recurrence_existence_1}
     r_{S}\left(1-\frac{S^*}{K_S}\right) = f_{S}\left(\frac{g_x+a_{x,S}S^*}{\mu_x}\right)G_{S}(N)h_{S}(S^*).
\end{equation}
Note that the LHS of Eq.~\eqref{eq:steady_recurrence_existence_1} is linear decreasing, taking the values in the interval $[0,r_2]$ in the invariant set $\mathcal{D}$, where $0$ is achieved for $S^*=K_S$. Let us denote the RHS of Eq.~\eqref{eq:steady_recurrence_existence_1} by $F$, that is, 
\[ F(Z)= G_{S}(N) f_{S}\left(\frac{g_x+a_{x,S}Z}{\mu_x}\right)h_{S}(Z), \quad Z \in [0,K_S]. \].

Let us fix $N\geq 0$ and consider the number of solutions to Eq.~\eqref{eq:steady_recurrence_existence_1}. It is obvious that for $N=0$, the only solution is $S^*=K_S$. 

For $N>0$, the function $F$ takes positive values for $Z\geq 0$. Now, we study the properties of $F$. Using Assumption~\ref{assumption1} we calculate:
\begin{itemize}
\item $F(0) = G_{S}(N)f_{S}\left(\frac{g_x}{\mu_x}\right)\in \left(G_S(N)a_{S,x}, G_S(N) \right] $; 
\item $F'(Z) = G_S(N) \left(\frac{a_{x,S}}{\mu_x} f'_S \left(\frac{g_x+a_{x,S}Z}{\mu_x}\right)h_S(Z)+ f_S\left(\frac{g_x+a_{x,S}Z}{\mu_x}\right)h_S'(Z) \right)\leq 0$;
\item 
$F(K_S) = G_{S}(N)f_{S}\left(\frac{g_x+a_{x,S}K_S}{\mu_x}\right)h_S(K_S)>0$,
$\lim\limits_{Z\to\infty} F(Z) =0$.
\end{itemize}
These properties allow us to conclude the following.
\begin{rem}
If $r_S> F(0)$, then there exists at least one RSS.
\end{rem}
However, without specifying the shapes of $f_S$ and $h_S$ more precisely, we are not able to estimate this number from above. Hence, we pose additional assumptions related to these functions. 
\begin{assumption}\label{assumption:convex}
Assume that $f_S$ and $h_S$ are convex functions.
\end{assumption}
\begin{rem}
The product of two nonnegative, decreasing, convex functions is a nonnegative, decreasing, convex function.
\end{rem}
Using this property of the function $F$, we are now in a position to
study the number of RSS.

\begin{itemize}
\item If $F(0)<r_S$, then there is exactly one RSS. The smaller the value of $F(0)$, the closer the solution $S^*$ is to $K_S$.
\item If $F(0)=r_S$, then:
\subitem (I) $|F'(0)| \leq \frac{r_S}{K_S}$, then there is only one solution that overlaps with the CCS.
\subitem (II) $|F'(0)| > \frac{r_S}{K_S}$, then there are two solutions and the smaller one overlaps with the CCS. 
\item If $F(0)>r_S$, then:
\subitem (I) $|F'(0)| \leq \frac{r_S}{K_S}$, then there is no nonnegative solution. 
\subitem (II) $|F'(0)| > \frac{r_S}{K_S}$, then there are two solutions until the next bifurcation, where $F(0)$ is such that the two solutions overlap and the straight line in tangent to $F$. For larger $F(0)$, there is no solution. 
\end{itemize}

One the basis of the above considerations, let us define the following critical values of $N$:
\[
\begin{split}
N_{th}^I  := & \ G_S^{-1}\left( \frac{r_S}{f_S(\frac{g_x}{\mu_x})} \right)=N_{\min}^{(S)}, \\
N_{th}^{II}  := & \ G_S^{-1}\left( 
\frac{r_S}
{K_S\left|
\frac{a_{x,S}}{\mu_x}f_S'(\frac{g_x}{\mu_x})+ f_S(\frac{g_x}{\mu_x})h_S'(0)\right|} 
\right),\\
N_{th}^{III} := & \ G_S^{-1}\left( \frac{r_S\left(1-\frac{S^*_{th}}{K_S} \right)}{f_S\left(\frac{g_x+a_{x,S}S^*_{th}}{\mu_x}\right)h_S(S^*_{th})} \right), \  \text{if there exists} \ S^*_{th} \ \text{satisfying}
\end{split}
\]
\begin{align*}
    \  \frac{\left(K_S-{S^*_{th}} \right)}{f_S\left(\frac{g_x+a_{x,S}S^*_{th}}{\mu_x}\right)h_S(S^*_{th})} 
    &= \frac{1}
    {\left|
    \frac{a_{x,S}}{\mu_x}f_S'(\frac{g_x+a_{x,S}S^*_{th}}{\mu_x})h_S(S^*_{th})+ f_S(\frac{g_x+a_{x,S}S^*_{th}}{\mu_x})h_S'(S^*_{th})\right|} . 
\end{align*}
Now, we are ready to formulate the following theorem.
\begin{theorem}
For System~\eqref{eq:our_system}, under Assumption~\ref{assumption:convex}, the number of the RSS varies from 0 to 2.
\begin{enumerate}
\item If $N<N_{th}^I$, then there is exactly one RSS.
\item At $N=N_{th}^I$ there is a bifurcation.
\item If $N^I_{th}< N_{th}^{II}$, then for $N \in \left( N_{th}^I,N_{th}^{II}\right]$ there is no RSS.
\item If $N^I_{th}> N_{th}^{II}$, then for $N > N_{th}^I$ there are two RSS until the next bifurcation at $N=N^{III}_{th}$.
\item If $N>N_{th}^{III}$, then there is no RSS.
\end{enumerate}
\end{theorem}

\begin{rem}
If we additionally assume $K_s>1$ (an obvious assumption in reality) and $h'(0)=-1$, then we can exclude Case 3.~in the above theorem due to the inequality $N_{th}^{II}<N_{th}^I$.
\end{rem}

\medskip

To gain a bit deeper insight into the existence and stability of the RSS and coexistence steady states, and to perform numerical simulations, in the following we introduce an additional assumption about the explicit forms of functions appearing in System~\eqref{eq:our_system}.
\begin{assumption}\label{assumption:explicit_functions} From now on we assume that
\begin{enumerate}
    \item $f_T(x)=\frac{a_{T,x} x + 1}{x+1}$ (decreasing if $a_{T,x}\in(0,1),$ $f_T(0)=1,$ $\lim_{x\rightarrow \infty}f_T(x)=a_{T,x}$),
    \item $g_T(u)=\frac{a_{T,u} u}{u+1}$ (increasing, $\lim_{u\rightarrow \infty}g_T(u)=a_{T,u}$),
    \item $h_T(T)=\frac{1}{T+1}$ (decreasing, $h_T(0)=1$, $\lim_{T\rightarrow \infty}h_T(T)=0$),
    \item $f_S(x)=\frac{a_{S,x} x + 1}{x+1}$ (decreasing if $a_{S,x}\in(0,1),$ $f_S(0)=1,$ $\lim_{x\rightarrow \infty}f_S(x)=a_{S,x}$),
    \item $g_S(u)=\frac{a_{S,u} u}{u+1}$ (increasing, $\lim_{u\rightarrow \infty}g_S(u)=a_{S,u}$),
    \item $h_S(S)=\frac{1}{S+1}$ (decreasing, $h_S(0)=1$, $\lim_{T\rightarrow \infty}h_S(S)=0$),
    \item $f_C(z)=\frac{a_{C,v} z}{z + 1}$ (increasing, $\lim_{z\rightarrow \infty}f_C(z)=a_{C,v}$, $f_C'(0)>0$, $\lim_{z\rightarrow \infty}f_C'(z)=0,$ where $z=(T+S)v$),
    \item $g_C(x)=\frac{a_{C,x} x + 1}{x+1}$ (decreasing for $a_{C,x}\in(0,1),$ $g_C(0)=1$, $\lim_{x\rightarrow \infty}g_C(x)$),
    \item $f_u(y)=\frac{a_{u,y} y}{y+1}+g_u$ (increasing, $f_u(0)=g_u,$ $\lim_{y\rightarrow \infty}f_u(z)=a_{u,y}$),
    \item $f_v(x)=\frac{1}{x+1}$ (decreasing, $f_v(0)=1,$ $\lim_{x\rightarrow \infty}f_v(x)=0$),
    \item $g_v(y)=\frac{a_{v,y} y}{y+1}$ (increasing, $g_v(0)=0,$ $\lim_{y\rightarrow \infty}g_v(y)=a_{v,y},$ $\lim_{y\rightarrow \infty}g_v'(y)=0$).
\end{enumerate}
\end{assumption}

We now reformulate conditions for the existence of RSS under Assumptions~\ref{assumption:explicit_functions}.
Since we assume that $T^*=0,$ then Eq.~\eqref{eq:our_system_steady_1} reads
\begin{equation}\label{eq:steady_recurrence_existence_11}
    \begin{aligned}
     0 &= R_{2}(S^*)S^* - f_{S}(x^*)G_{2}(N)h_{S}(S^*)S^*.
    \end{aligned}
\end{equation}

\begin{figure}[t]
    \centering
    
    \begin{subfigure}{0.49\textwidth}
    \centering
    \captionsetup{justification=centering,margin=0.5cm}
\includegraphics[width=1\textwidth]{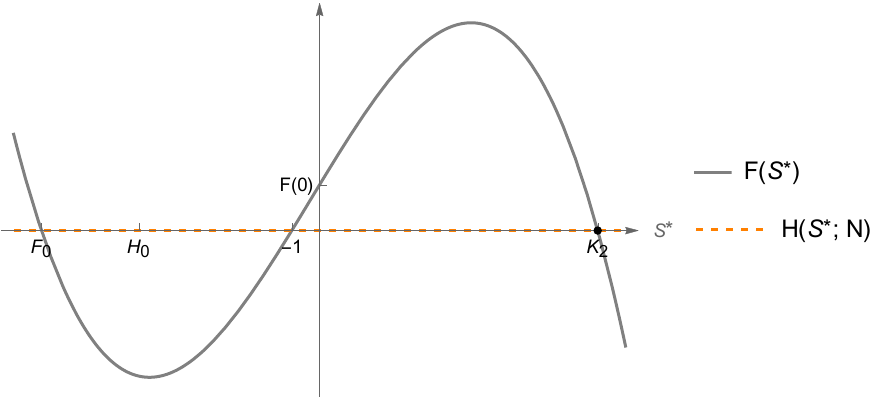}
    \caption{$N = 0$}
    \label{fig:0Sx_N=0}
    \end{subfigure}
    \hfill
    \begin{subfigure}{0.49\textwidth}
    \centering
    \captionsetup{justification=centering,margin=0.5cm}   \includegraphics[width=1\textwidth]{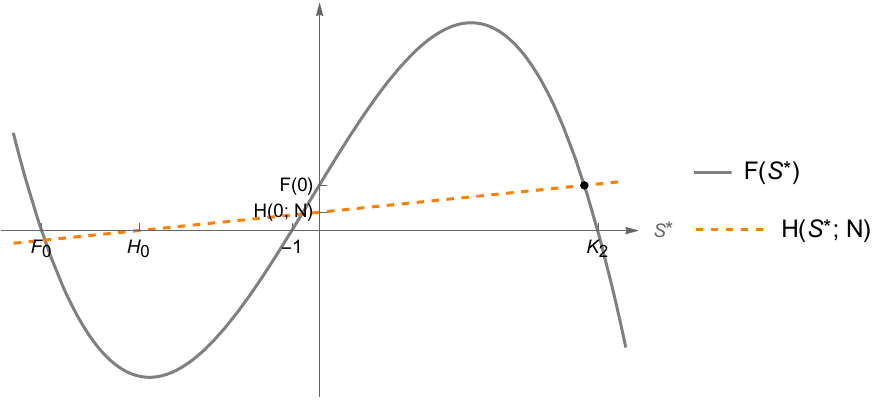}
    \caption{$H(0;N)<F(0)$}
    \label{fig:0Sx_H0_less_F0}
    \end{subfigure}
    \begin{subfigure}{0.49\textwidth}
    \centering
    \captionsetup{justification=centering,margin=0.5cm}
\includegraphics[width=1\textwidth]{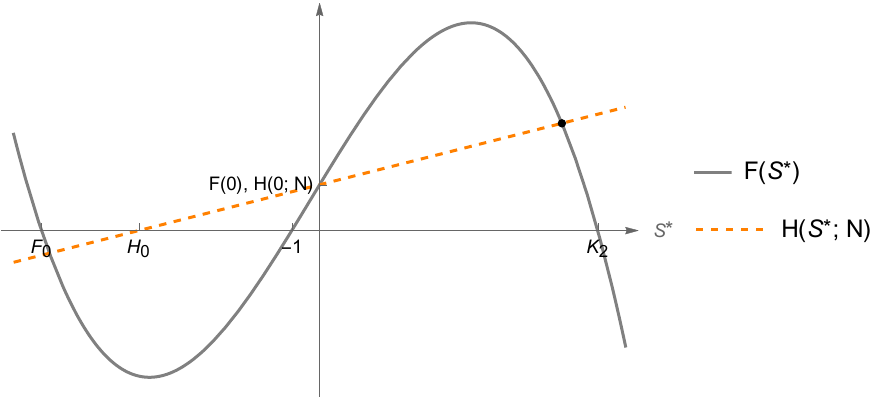}
    \caption{$H(0; N)=F(0),$ $H'(0;N)<F'(0)$}
    \label{fig:0Sx_H0_eq_F0}
    \end{subfigure}
    \hfill
    \begin{subfigure}{0.49\textwidth}
    \centering
    \captionsetup{justification=centering,margin=0.5cm}   \includegraphics[width=1\textwidth]{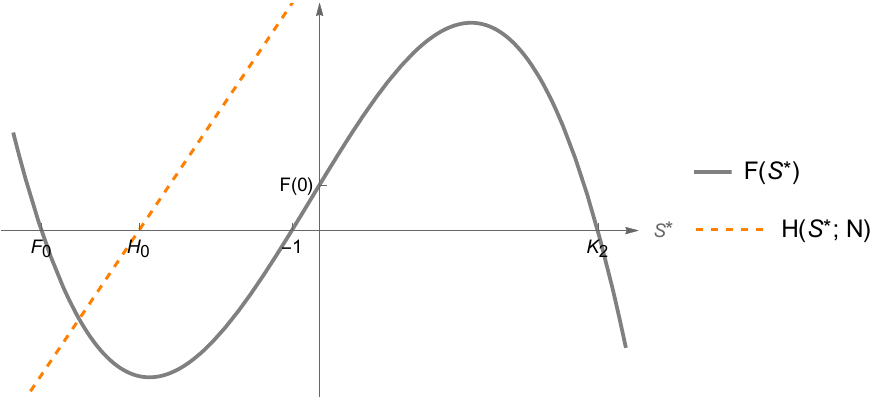}
    \caption{$H(0; N)\geqslant F(0), H'(0; N)\geqslant  F'(0)$}
    \label{fig:0Sx_H0_more_more_F0}
    \end{subfigure}
    
    \caption{Sample plots of $H(S^*;N)$ and $F(S^*)$ depending on the value of the parameter $N$.}
    \label{fig:0Sx_1}
\end{figure}

We divide both sides of Eq.~\eqref{eq:steady_recurrence_existence_11} by $S^*>0$ and substitute Eq.~\eqref{eq:x_steady} to obtain 

\begin{equation}\label{eq:steady_recurrence_existence_2}
    \begin{aligned}
    0 &=r_S \frac{K_S-S^*}{K_S} - \frac{a_{S,x}(g_x+a_{x,S}S^*)+\mu_x}{g_x+a_{x,S}S^*+\mu_x}\frac{1}{S^*+1}G_{S}(N).
    \end{aligned}
\end{equation}

By reducing the RHS of Eq.~\eqref{eq:steady_recurrence_existence_2} to the common denominator and taking only the numerator we get the equivalent equation that reads
\begin{equation}\label{eq:steady_recurrence_existence_3}
    \begin{aligned}
    r_S(K_S-S^*)(g_x+a_{x,S}S^*+\mu_x)(S^*+1)=K_S(a_{S,x}(g_x+a_{x,S}S^*) + \mu_x)G_S(N).
    \end{aligned}
\end{equation}
Let $F(S^*),$ $H(S^*;N)$ denote the LHS and RHS of Eq.~\eqref{eq:steady_recurrence_existence_3}, respectively. Of course, $F(S^*)$ is a third degree polynomial of $S^*,$ while $H(S^*;N)$ is a first-degree polynomial of $S^*$ with $N$ as a parameter. Note that finding the number of RSSs depending on the value of $N$ is equivalent to finding the number of points at which the graphs of $F$ and $H$ intersect for $S^*>0$ depending on the value of $N$.
The set of roots of $F(S^*)$ reads 
\begin{align*}
    \mathcal{F}^0=\left\{-\frac{g_x + \mu_x}{a_{x,S}}, -1, K_S\right\},
\end{align*}
while the only root of $H(S^*;N)$ is equal to
$$-\frac{\mu_x + g_x a_{S,x}}{a_{x,S}a_{S,x}}.$$
We easily see that
\begin{subequations}\label{eq:F_H_0}
\renewcommand{\theequation}{\theparentequation.\arabic{equation}}
  \begin{align}
    F(0)=&\; r_S K_S (g_x + \mu_x), \label{eq:F_0}\\
    F'(0)=&\;r_S \Big((g_x+\mu_x)(K_S - 1) + K_S a_{x,S}\Big),\label{eq:F'_0}\\
    H(0;N)=&\;G_S(N) K_S (a_{S,x} g_x + \mu_x),\label{eq:H_0}\\
    H'(0;N)=&\;G_S(N) K_S a_{S,x} a_{x,S}.\label{eq:H'}
  \end{align}
\end{subequations}
As we have noticed before in the general case, in reality we have $K_S\gg 1$, and therefore $F'(0)>0$. This implies that the third-degree polynomial $F$ is concave for $S^*>0$, since $F(0)$ is positive.

\begin{figure}[t]
    \centering
    
    \begin{subfigure}{0.49\textwidth}
    \centering
\includegraphics[width=1\textwidth]{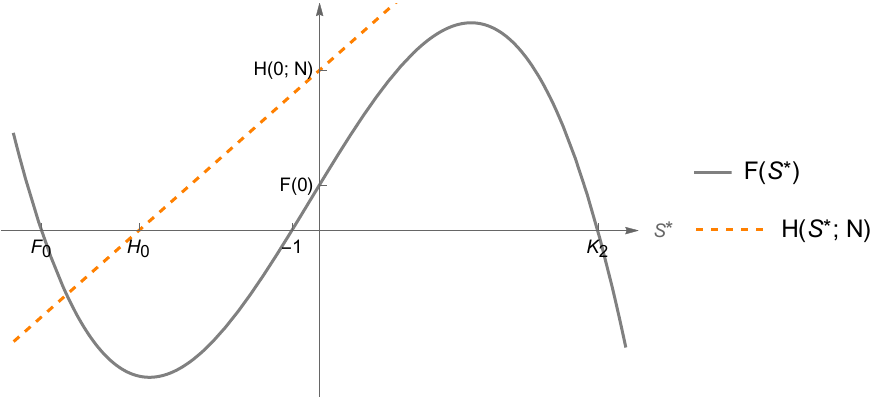}
    \caption{$H(0)> F(0)$, $H' < F'(0), N>N_{\text{crit}}$}
    \label{fig:0Sx_N_greater_Ncrit}
    \end{subfigure}
    \hfill
    \begin{subfigure}{0.49\textwidth}
    \centering
\includegraphics[width=1\textwidth]{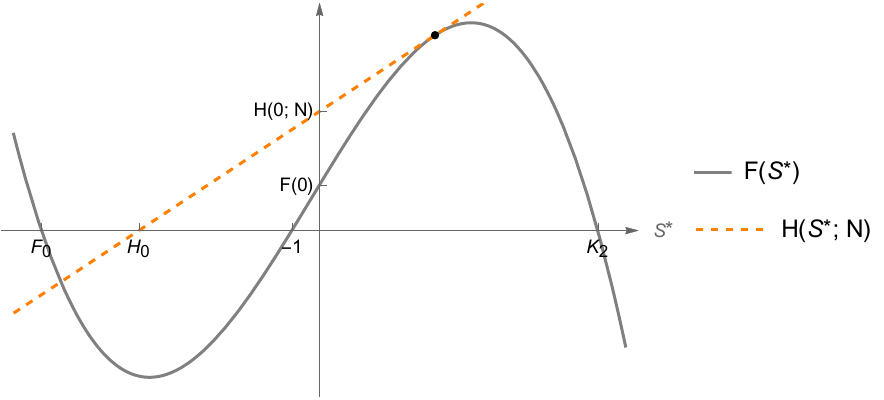}
    \caption{$H(0)> F(0)$, $H' < F'(0), N=N_{\text{crit}}$}
    \label{fig:0Sx_N_eq_Ncrit}
    \end{subfigure}
    
    \begin{subfigure}{0.49\textwidth}
    \centering
\includegraphics[width=1\textwidth]{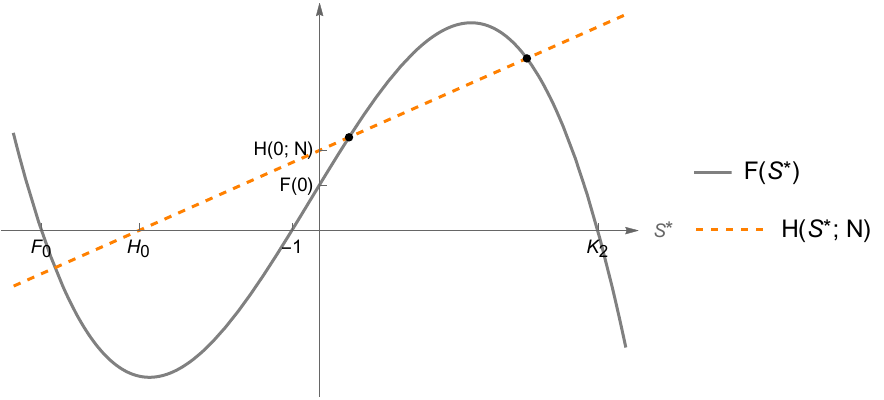}
    \caption{$H(0)> F(0)$, $H' < F'(0), N<N_{\text{crit}}$}
    \label{fig:0Sx_N_less_Ncrit}
    \end{subfigure}
    
    \caption{Sample plots of $H(S^*;N)$ and $F(S^*)$ depending on the value of the parameter $N;$ illustration of a bifurcation when the number of steady states changes from $0,$ through $1,$ to $2.$}
    \label{fig:0Sx_11}
\end{figure}

In this case, if $H(0;N)<F(0),$ then the graphs of $F$ and $H$ will intersect exactly once for $S^*>0,$ see Figs~\ref{fig:0Sx_N=0},~\ref{fig:0Sx_H0_less_F0}, as well as if $H(0;N)=F(0)$ and $H'(0;N)<F'(0),$ see Fig.~\ref{fig:0Sx_H0_eq_F0}. However, if $H(0;N)\geqslant F(0)$ and $H'(0;N)\geqslant F'(0),$ then the graphs of $F$ and $H$ will not intersect for $S^*>0,$ see Fig.~\ref{fig:0Sx_H0_more_more_F0}. 

If conditions $H(0;N)>F(0)$ and $H'(0;N)<F'(0)$ are met for some nonnegative $N_0$, then there exists an open interval $I$ such that $N_0\in I$ and $H(0;N)>F(0),$ $H'(0;N)<F'(0)$ for all $N\in I.$ Furthermore, there exists $N_{\text{crit}}\in I$ for which the graph of $H$ is tangent to the graph of $F$ at some $S^{*}_{\text{crit}},$ i.e.~the graphs of $H$ and $F$ intersect at exactly one point for $S^*>0,$ see Fig.~\ref{fig:0Sx_N_eq_Ncrit}. Furthermore, if $N\in I$ and $N<N_{\text{crit}},$ then the graphs of $H$ and $F$ intersect twice for $S^*>0,$ see Fig.~\ref{fig:0Sx_N_less_Ncrit}. On the other hand, if $N\in I$ and $N>N_{\text{crit}},$ then the graphs of $H$ and $F$ do not intersect for $S^*>0,$ see Fig.~\ref{fig:0Sx_N_greater_Ncrit}.

In order to find $N_{\text{crit}}$ and $S^{*}_{\text{crit}}$ let us solve the system
\begin{equation*}
    \begin{aligned}
    H(T^*;N)&=F(T^*),\\
    H'(T^*;N)&=F'(T^*),
    \end{aligned}
\end{equation*}
which is equivalent to
\begin{subequations} \label{eq:N_crit_S_crit}
\renewcommand{\theequation}{\theparentequation.\arabic{equation}}
  \begin{align}
    K_S\big(a_{S,x} (g_x + a_{x,S}S^*) + \mu_x\big) G_S(N)&=F(S^*),\label{eq:N_crit_S_crit_1}\\
    K_S a_{S,x}a_{x,S}G_S(N)&=F'(S^*).\label{eq:N_crit_S_crit_2}
  \end{align}
\end{subequations}
From Eq.~\eqref{eq:N_crit_S_crit_2} we get
\begin{equation}\label{eq:N_crit_1}
    \begin{gathered}
    G_S(N)=\frac{F'(S^*)}{K_S a_{S,x} a_{x,S}}
    \end{gathered}
\end{equation}
and after substituting Eq.~\eqref{eq:N_crit_1} into Eq.~\eqref{eq:N_crit_S_crit_1} we obtain
\begin{equation}\label{eq:N_crit_S_crit_3}
\begin{gathered}
\frac{F'(S^*)}{a_{S,x} a_{x,S}}\Big(a_{S,x}(g_x+a_{x,S}S^*) + \mu_x\Big)-F(S^*)=0.
\end{gathered}
\end{equation}
Of course, the LHS of Eq.~\eqref{eq:N_crit_S_crit_3} is a third-degree polynomial so we can obtain its roots analytically. Let us denote them as $w_0,w_1,w_2.$ Since we know that there exists exactly one positive $S^{*}$ that solves Eq.~\eqref{eq:N_crit_S_crit}, then we will obtain $S^{*}_{\text{crit}}$ by taking
\begin{align}\label{eq:S_crit}
    S^{*}_{\text{crit}} = \max \left\{ \Re(w_0),\Re(w_1), \Re(w_2) \right\}.
\end{align}
On the other hand,  applying $G_{S}^{-1}$ to both sides of Eq.~\eqref{eq:N_crit_1} we obtain
\begin{align*}\label{eq:N_crit_2}
    N_{\text{crit}}&=G_{S}^{-1}\left( \frac{F'(S^{*}_{\text{crit}})}{ K_S a_{S,x}a_{x,S}}\right).
\end{align*}

\begin{cor}\label{cor:0Sx_steady_existence}
    The number of the RSS of System~\eqref{eq:our_system} under Assumption~\ref{assumption:explicit_functions} is dependent on the value of the parameter $N$ in the following way. For $N\geqslant 0$ such that
    \begin{enumerate}
        \item $H(0;N) \geqslant F(0)$ and $H'(0;N) \geqslant F'(0)$ there are no RSS;
        \item $H(0;N) < F(0)$ or $H(0;N) = F(0)$ and $H'(0;N) < F'(0)$ there is exactly one RSS;
        \item $H(0;N) > F(0)$ and $H'(0;N) < F(0)$ and
            \begin{enumerate}
                \item $N>N_{\text{crit}}$ there are no RSS;
                \item $N=N_{\text{crit}}$ there is exactly one RSS;
                \item $N<N_{\text{crit}}$ there are two RSS.
            \end{enumerate}
    \end{enumerate}
\end{cor}

Let us now proceed to analyze the RSS stability. From Eq.~\eqref{jacobi} we have

\begin{equation}\label{eq:jacobi_0Sx}
    \begin{aligned}
    \mathcal{J}(0,S^*,x^*) &= 
    \begin{pmatrix}
    \frac{\partial f_1}{\partial T}(0,S^*,x^*) & 0 & 0
    \\[2ex]
    \frac{\partial f_2}{\partial T}(0,S^*,x^*) & \frac{\partial f_2}{\partial S}(0,S^*,x^*) & \frac{\partial f_2}{\partial x}(0,S^*,x^*)
    \\[2ex]
    a_{x,T} & a_{x,S} & -\mu_{x}
    \end{pmatrix}
    \end{aligned}
\end{equation}
The characteristic polynomial of $\mathcal{J}(0,S^*,x^*)$ is given as
\begin{equation*}
    \begin{aligned}
    W(\lambda)&=
    \left(\frac{\partial f_1}{\partial T}-\lambda\right)\left( \lambda^2 +\lambda \left(\mu_{x}-\frac{\partial f_2}{\partial S}\right)
    -\left(\mu_{x}\frac{\partial f_2}{\partial S} +  a_{x,S}\frac{\partial f_2}{\partial x} \right)\right).
    \end{aligned}  
\end{equation*}
Let us denote
\begin{equation}\label{RSSstab}
    \begin{gathered}
    V(\lambda):=\lambda^2 +\lambda \left(\mu_{x}-\frac{\partial f_2}{\partial S}\right)
    -\left(\mu_{x}\frac{\partial f_2}{\partial S} + a_{x,S}\frac{\partial f_2}{\partial x} \right)
    \end{gathered}
\end{equation}
and let $v_1, v_2$ be the roots of $V(\lambda).$ Of course, if all the eigenvalues of $\mathcal{J}(0,S^*,x^*),$ i.e.~$\frac{\partial f_1}{\partial T}, v_1, v_2,$ have negative real parts, then the RSS is locally asymptotically stable. Note that from Eq.~\eqref{eq:x_steady} and Assumption~\ref{assumption:explicit_functions} we obtain
\begin{equation}\label{eq:df1/dT_0Sx}
    \begin{aligned}
    \frac{\partial f_1}{\partial T}(0,S^*,x^*)&=\frac{r_{\alpha}S^*}{K_S}+r_T - \frac{a_{T,x}(g_x+a_{x,S}S^*)+\mu_x}{g_x+a_{x,S}S^*+\mu_x}G_T(N) \\
    & = \frac{r_{\alpha}S^*}{K_S}+r_T - 
    \left(a_{T,x}+
\frac{(1-a_{T,x})\mu_x}{g_x+a_{x,S}S^*+\mu_x}\right)
    G_T(N) .
    \end{aligned}
\end{equation}
It is easy to see that for the values of $N$ close to $0$ the negative component of the RHS of Eq.~\eqref{eq:df1/dT_0Sx} is also close to $0$ and thus $\frac{\partial f_1}{\partial T}>0.$ 
On the other hand, assuming $a_{T,x}\leq 1$ and remembering that $S^*\leq K_S$ we see that it is enough to require $G_T(N)> \frac{r_{\alpha}+r_T}{a_{T,x}}$ to get 
$\frac{\partial f_1}{\partial T}< 0.$ 

Note also that for the RSS we can determine the value of $N$ as a function of $S^*.$ From the condition $H(S^*;N)=F(S^*)$ we get
\begin{equation*}
    \begin{gathered}
    K_S(a_{S,x}(g_x+a_{x,S}S^*)+\mu_x)G_S(N)=F(S^*),\\
    \end{gathered}
\end{equation*}
or equivalently 
\begin{equation}\label{eq:0Sx_N_depending _on_S}
    \begin{aligned}
    N&=G_S^{-1}\left( \frac{F(S^*)}{K_S(a_{S,x}(g_x+a_{x,S}S^*)+\mu_x)}\right).
    \end{aligned}
\end{equation}

From Corollary~\ref{cor:G1_G2} and Eq.~\eqref{eq:df1/dT_0Sx} we can deduce that

\begin{equation}\label{eq:df1/dT_leq_1}
    \begin{aligned}
        \frac{\partial f_1}{\partial T}(0,S^*,x^*) 
        & \leqslant \frac{r_{\alpha}S^*}{K_S}+r_T - \frac{a_{T,x}(g_x+a_{x,S}S^*)+\mu_x}{g_x+a_{x,S}S^*+\mu_x}G_S\left(N\right).
    \end{aligned}
\end{equation}

\begin{assumption}\label{assumption:atx=asx}
    From now on, we assume that $a_{T,x}=a_{S,x}.$
\end{assumption}

From the Assumption above and Eq.~\eqref{eq:0Sx_N_depending _on_S} we get

\begin{equation}
    \begin{aligned}
        \frac{a_{T,x}(g_x+a_{x,S}S^*)+\mu_x}{g_x+a_{x,S}S^*+\mu_x}G_S(N) &= \frac{r_S (K_S - S^*)(S^*+1)}{K_S},
    \end{aligned}
\end{equation}

which implies that

\begin{equation}\label{eq:df1/dT_leq_2}
    \begin{aligned}
        \frac{\partial f_1}{\partial T}(0,S^*,x^*) 
        & \leqslant \frac{r_{\alpha}S^* + r_T K_S - r_S (K_S - S^*)(S^*+1)}{K_S}.
    \end{aligned}
\end{equation}

Let us denote the numerator of the RHS of the inequality above as

\begin{equation*}
    \begin{aligned}
         Q(S^*)&:=r_{\alpha}S^*+r_T K_S - r_S (K_S - S^*)(S^*+1)
         \\
         & \ =r_S S^{*2} + S^*(r_S - r_S K_S + r_{\alpha}) + K_S(r_T-r_S).
     \end{aligned}
\end{equation*}

Then we have

\begin{equation}
    \begin{aligned}
        Q(S^*)<0 \implies \frac{\partial f_1}{\partial T}(0,S^*,x^*)<0.
    \end{aligned}
\end{equation}

Notice that the above condition can be satisfied only for small values of $S^*$, since for $S^*$ near $K_S$ the positive terms of $Q(S^*)$ are much larger compared to the negative one.
Coming back to Formula~\eqref{RSSstab}, to get  stability of the RSS it is enough to assume

\begin{equation}\label{eq:v1_v2_negative}
    \begin{aligned}
    \frac{\partial f_2}{\partial S} - \mu_x&<0 \quad &\land \quad
    -\mu_x \frac{\partial f_2}{\partial S} - a_{S,x}\frac{\partial f_2}{\partial x} &> 0.
    \end{aligned}
\end{equation}

We have

\begin{equation*}
    \begin{aligned}
        \frac{\partial f_2}{\partial S} - \mu_x &= 
        r_S \Big(1-\frac{2S^*}{K_S}\Big) - G_S(N)f_S(x^*)\frac{1}{(S^*+1)^2} - \mu_x,
    \end{aligned}
\end{equation*}

and note that to have the first inequality in \eqref{eq:v1_v2_negative} satisfied it is enough to assume $r_S\leq \mu_x$. In the more general case, substituting $G_S(N)$ we obtain

\begin{equation*}
    \begin{aligned}
        \frac{\partial f_2}{\partial S} - \mu_x &= 
        r_S \Big(1-\frac{2S^*}{K_S}\Big) -\frac{F(S^*)}{K_S(g_x + a_{x,T}T^* + a_{x,S}S^*+\mu_x)(S^*+1)^2} - \mu_x
        \\
        &= \frac{r_S (K_S - 2S^*)(S^*+1) - r_S(K_S-S^*) - \mu_x K_S (S^*+1)}{K_S(S^*+1)}.
    \end{aligned}
\end{equation*}

Let us denote the numerator of the expression above as

\begin{equation*}
    \begin{aligned}
        P(S^*)&:=r_S (K_S - 2S^*)(S^*+1) - r_S(K_S-S^*) - \mu_x K_S (S^*+1)\\
        & \ = -2r_S{S^*}^2 + S^*\big(r_S(K_S - 1) - \mu_x K_S\big) - \mu_x K_S.
    \end{aligned}
\end{equation*}

Then

\begin{equation*}
    \begin{gathered}
         \frac{\partial f_2}{\partial S} - \mu_x <0 \iff 
         P(S^*) <0.
    \end{gathered}
\end{equation*}

It is obvious that $P(S^*)<0$ for the values of $S^*$ exceeding $\frac{K_S}{2}$.
Moreover, since

\begin{equation*}
    \begin{aligned}
        \frac{\partial f_2}{\partial x}&=-f_S'(x^*)G_S(N)h_S(S^*)S^*=-\frac{a_{S,x}-1}{(x^*+1)^2}G_S(N) \frac{S^*}{S^*+1}
        \\
        &=\frac{\mu_x r_S(1-a_{S,x})S^*(K_S-S^*)}{K_S (g_x + a_{x,S}S^*+\mu_x)(a_{S,x}(g_x+a_{x,S}S^*)+\mu_x)},
    \end{aligned}
\end{equation*}

then

\begin{equation*}
    \begin{aligned}
        -\mu_x \frac{\partial f_2}{\partial S} - a_{S,x}\frac{\partial f_2}{\partial x} =& 
        \mu_x\frac{r_S(K_S-S^*) -r_S (K_S - 2S^*)(S^*+1)}{K_S(S^*+1)} \\
        &+a_{S,x}\frac{\mu_x r_S(a_{S,x}-1)S^*(K_S-S^*)}{K_S (g_x + a_{x,S}S^*+\mu_x)(a_{S,x}(g_x+a_{x,S}S^*)+\mu_x)}.
    \end{aligned}
\end{equation*}

After reducing the RHS of the equation above to a common denominator, we can denote its numerator as

\begin{equation*}
    \begin{aligned}
        B(S^*):=&S^*(2{S^*}+1-K_S)(g_x + a_{x,S}S^*+\mu_x)\big(a_{S,x}(g_x+a_{x,S}S^*)+\mu_x\big)\\ &-a_{S,x}(1-a_{S,x})S^*(K_S-S^*)(S^*+1).
    \end{aligned}
\end{equation*}

Thus,

\begin{equation*}
    \begin{gathered}
        -\mu_x \frac{\partial f_2}{\partial S} - a_{S,x}\frac{\partial f_2}{\partial x}>0 \iff B(S^*)>0.
    \end{gathered}
\end{equation*}

Note that $B(S^*)<0$ for small values of $S^*$.

\begin{cor}
    The RSS of System~\eqref{eq:our_system}  is locally asymptotically stable if $Q(S^*)<0$, $P(S^*)<0$, and $B(S^*)>0.$
\end{cor}

\vspace*{0.3cm}
\subsection{Coexistence steady states}

Now we will consider a situation in which a certain population of both types of cancer cells survives, i.e.~we will analyze the existence and stability of positive steady states $(T^*,S^*,x^*).$ To be consistent with the previous subsections, we will call this state the PSS. 

Recall that if $N=0$, then  System \eqref{eq:our_system_steady} simplifies to~\eqref{N=0} and the only positive 
solution to it is
\begin{equation}\label{steady_K1_K2}
    \big(K_T,K_S,x^*_{K_T,K_S}\big)=\Big(K_T,K_S,\frac{g_x + a_{x,T}K_T + a_{x,S}K_S}{\mu_x}\Big),
\end{equation}
which is a stable node.
Since System \eqref{eq:our_system} depends continuously on $N$, for the values of $N$ sufficiently close to $0$ there should also exist a locally asymptotically stable steady state $(T^*,S^*,x^*)$ with $T^*,S^*>0$. Moreover, the coordinates $T^*$, $S^*$ remain close to $K_T$ and $K_S$, respectively.

We will now state the conditions necessary for the point $(S^*, T^*, x^*),$ $S^*,T^*>0$ to be the solution to System \eqref{eq:our_system_steady}. Adding the first two equations of System~\eqref{eq:our_system_steady} by sides we get
\begin{equation}\label{eq:1i2_dodane}
    \begin{gathered}
    - r_T(T^*)T^* + f_T(x^*)G_T(N)h_T(T^*)T^* = r_S(S^*)S^* - f_S(x^*)G_S(N)h_S(S^*)S^*.
    \end{gathered}
\end{equation}
In particular, Equation \eqref{eq:1i2_dodane} holds when both sides are equal to $0$, that is
\begin{equation}\label{eq:warunek0}
    \begin{aligned}
    G_T(N)&=\frac{r_T(K_T-T^*)}{K_T f_T(x^*)h_T(T^*)}, &
    G_S(N)&=\frac{r_S(K_S-S^*)}{K_S f_S(x^*)h_S(S^*)}.
    \end{aligned}
\end{equation}
By substituting \eqref{eq:warunek0} into System \eqref{eq:our_system_steady} we get
\begin{equation*}
    \begin{aligned}
    0&=\frac{r_{\alpha}T^*S^*}{K_T K_S}(K_T-T^*),\\
    0&=-\frac{r_{\alpha}T^*S^*}{K_T K_S}(K_T-T^*),\\
    0&=g_x + a_{x,T}T^* + a_{x,S}S^*-\mu_x x^*.
    \end{aligned}
\end{equation*}
Since $T^*, S^*>0$, solutions to the system above must have the form
\begin{equation}\label{steady_K1_S}
    \big(T^*,S^*,x^*\big)=\Big(K_T,S^*,\frac{g_x + a_{x,T}K_T + a_{x,S}S^*}{\mu_x}\Big),
\end{equation}
where $S^*>0$.
Substituting \eqref{steady_K1_S} into System \eqref{eq:warunek0} we get
\begin{equation}\label{eq:warunek01}
    \begin{aligned}
    G_T(N)&=0, &
    G_S(N)&=\frac{r_S(K_S-S^*)}{K_S f_S(x^*)h_S(S^*)}.
    \end{aligned}
\end{equation}
From Corollary \ref{cor:G1_G2} we conclude that Condition \eqref{eq:warunek01} holds if and only if $N=0,$ $S^*=0$, but we have already analyzed this case.

Thus, let us assume that both sides of Equation~\eqref{eq:1i2_dodane} are different from $0$. In this case, both sides need to be of the same sign. Both sides of this equation are positive if and only if
\begin{equation*}
    \begin{aligned}
    G_T(N)&>\frac{r_T(K_T-T^*)}{K_T f_T(x^*)h_T(T^*)}, &
    G_S(N)&<\frac{r_S(K_S-S^*)}{K_S f_S(x^*)h_S(S^*)},
    \end{aligned}
\end{equation*}
and both sides are negative if and only if
\begin{equation}\label{eq:warunek2}
    \begin{aligned}
    G_T(N)&<\frac{r_T(K_T-T^*)}{K_T f_T(x^*)h_T(T^*)}, &
    G_S(N)&>\frac{r_S(K_S-S^*)}{K_S f_S(x^*)h_S(S^*)}.
    \end{aligned}
\end{equation}
Note that since $\forall{N> 0} \;G_T(N)>G_S(N),$ then Condition \eqref{eq:warunek2} does not hold for any $N$ if
\begin{equation*}
    \frac{r_S(K_S-S^*)}{K_S f_S(x^*)h_S(S^*)} \geqslant
    \frac{r_T(K_T-T^*)}{K_T f_T(x^*)h_T(T^*)}.
\end{equation*}
\begin{cor}\label{wn:konieczne_na_TSx}
Let $T^*\in(0,K_T]$, $S^*\in(0,K_T]$ and let Assumptions \ref{assumption1}, \ref{assumption2} and \ref{assumption:explicit_functions} be satisfied. If $(T^*,S^*,x^*)$ is a steady state of System \eqref{eq:our_system}, then one of the following conditions holds
\begin{enumerate}
    \item $N=0$,
    \item $ G_T(N)>\frac{r_T(K_T-T^*)}{K_T f_T(x^*)h_T(T^*)},
    G_S(N)<\frac{r_S(K_S-S^*)}{K_S f_S(x^*)h_S(S^*)},$
    \item $G_T(N)<\frac{r_T(K_T-T^*)}{K_T f_T(x^*)h_T(T^*)},
    G_S(N)>\frac{r_S(K_S-S^*)}{K_S f_S(x^*)h_S(S^*)}.$
\end{enumerate}
\end{cor}
It is easy to see that if $N$ satisfies
\begin{equation*}
    \begin{aligned}
    G_T(N)&>\frac{r_T(K_T-T^*)}{K_T f_T(x^*)h_T(T^*)}, &
    G_S(N)&>\frac{r_S(K_S-S^*)}{K_S f_S(x^*)h_S(S^*)},
    \end{aligned}
\end{equation*}
then $N$ does not satisfy any of the conditions from Corollary \ref{wn:konieczne_na_TSx}. Moreover, for any given $T^*\in(0,K_T]$, $S^*\in(0,K_S]$ we have
\begin{equation*}
    \begin{aligned}
    \frac{r_T(K_T-T^*)}{K_T f_T(x^*)h_T(T^*)}=&
    \frac{r_T(K_T-T^*)(g_x + a_{x,T}T^*+a_{x,S}S^*+\mu_x)(T^*+1)}{K_T \big(a_{T,x}(g_x + a_{x,T}T^*+a_{x,S}S^*)+\mu_x\big)}\\
    <&\frac{r_T(g_x+a_{x,T}K_T + a_{x,S}K_S+\mu_x)(K_T+1)}{a_{T,x}g_x+\mu_x},\\
    \frac{r_S(K_S-S^*)}{K_S f_S(x^*)h_S(S^*)}=&
     \frac{r_S(K_S-S^*)(g_x + a_{x,T}T^*+a_{x,S}S^*+\mu_x)(S^*+1)}{K_S \big(a_{S,x}(g_x + a_{x,T}T^*+a_{x,S}S^*)+\mu_x\big)}\\
     <&\frac{r_S(g_x+a_{x,T}K_T + a_{x,S}K_S+\mu_x)(K_S+1)}{a_{S,x}g_x+\mu_x},
    \end{aligned}
\end{equation*}
implying the following corollary.
\begin{cor}\label{wn:nie_ma_TSx}
Let Assumptions \ref{assumption1}, \ref{assumption2}, and \ref{assumption:explicit_functions} hold. If $N$ satisfies
\begin{equation*}
    G_i(N)\geqslant\frac{r_i(g_x+a_{x,T}K_T + a_{x,S}K_S+\mu_x)(K_i+1)}{a_{T,x}g_x+\mu_x},\ \ i=T,\, S,
\end{equation*}
then System \eqref{eq:our_system} has no steady states $(T^*,S^*,x^*)$, such that $T^*\in(0,K_T]$ and $S^*\in(0,K_S].$
\end{cor}

$\newline$

\subsection{Numerical simulations}

In this section, we would like to illustrate the behavior of System \eqref{eq:our_system} with numerical simulations performed in \textit{Wolfram Mathematica}. For this purpose, we will use fixed values of the model parameters (see Table \ref{table1}), selected initial conditions, and values of $N$. Moreover, we will need the explicit forms of the $G_T$ and $G_S$ functions. Under Assumption \ref{assumption:explicit_functions} we get

\begin{equation*}
    \begin{aligned}
    G_i(N)=&\,g_i\Big(\frac{1}{\mu_u}  f_u\Big(\frac{a_{y,C}N}{\mu_y \mu_C}\Big)\Big)\frac{N}{\mu_C}\\
    =&\,\frac{Na_{y,C}a_{i,u}(a_{u,y}+g_u)+g_u a_{i,u}\mu_y \mu_C}{Na_{y,C}(a_{u,y}+g_u+\mu_u)+\mu_y \mu_C(g_u +\mu_u)}\frac{N}{\mu_C},\ \ i=T,\, S.
    \end{aligned}
\end{equation*}

\begin{table}
\centering
\begin{tabularx}{0.8\textwidth}{|>{\centering\arraybackslash}X | >{\centering\arraybackslash}X | >{\centering\arraybackslash}X |}
    \hline
    Parameter & Value & Unit \\
    \hline
     $K_T$ & $10^8$ & cell  \\
     $K_S$ & $10^7$ & cell \\
     $r_T$ & $10^{-3}$ & h$^{-1}$  \\
     $r_S$ & $0,1$ & h$^{-1}$  \\
     $r_{\alpha}$ & $6 \cdot 10^{-3}$ & h$^{-1}$  \\
     $a_{T,x}$ & $0,69$ & -- \\
     $a_{S,x}$ & $0,69$ & -- \\
     $a_{T,u}$ & $0,12$ & h$^{-1}$  \\
     $a_{S,u}$ & $1,2\cdot 10^{-2}$ & h$^{-1}$  \\
     $a_{x,T}$ & $5,75\cdot 10^{-6}$ & pg$\cdot$cell$^{-1}$h$^{-1}$  \\
     $a_{x,S}$ & $5,75\cdot 10^{-6}$ & pg$\cdot$cell$^{-1}$h$^{-1}$  \\
     $a_{y,C}$ & $1,02\cdot 10^{-4}$ & pg$\cdot$cell$^{-1}$h$^{-1}$  \\
     $a_{u,y}$ & $2,88$ & rec$\cdot$cell$^{-1}$h$^{-1}$  \\
     $g_x$ & $6,3945\cdot 10^4$ & pg$\cdot$h$^{-1}$ \\
     $g_u$ & $1,44$ & rec$\cdot$cell$^{-1}$h$^{-1}$ \\
     $\mu_C$ & $7\cdot 10^{-3}$ & h$^{-1}$  \\
     $\mu_x$ & $7$ & h$^{-1}$  \\
     $\mu_y$ & $0,102$ & h$^{-1}$  \\
     $\mu_u$ & $1,44\cdot 10^{-2}$ & h$^{-1}$  \\
     \hline
\end{tabularx}
\caption{Values of parameters taken from \cite{Abernathy}, where pg, rec are picograms and receptors, respectively.}
\label{table1}
\end{table}

Having the explicit forms of all the functions appearing in System~\eqref{eq:our_system} we can now solve this system numerically for given initial conditions and values of $N$. Let us introduce auxiliary functions
\begin{align*}
    \bar{T}_i(N):=&T(t_i;T_{0i}, S_{0i}, x_{0i}, N),\\
    \bar{S}_i(N):=&S(t_i;T_{0i}, S_{0i}, x_{0i}, N),
\end{align*}
i.e.~$\bar{T}_i(N),$ $\bar{S}_i(N)$ are the values of the respective coordinates of the solution of System \eqref{eq:our_system} at $t$ equal to some $t_i,$ for some initial conditions $T_{0i}, S_{0i}, x_{0i}$ and for a given value of $N.$ In our model time is expressed in hours, so $t^*=2160,$ $t^{**}=26352$ correspond to approximately $3$ months and $3$ years, respectively.

Earlier we have shown, that if Condition \eqref{eq:G1_G2_triv} is satisfied, then the point $\left(0,0,\frac{g_x}{\mu_x}\right)$ is a locally asymptotically stable steady state of System \eqref{eq:our_system}:
\begin{equation*}
    \begin{aligned}
    G_T(N)>&\,\frac{r_T(g_x+\mu_x)}{a_{T,x}g_x}\approx 1.45\cdot10^{-3}, & G_S(N)>&\,\frac{r_S(g_x+\mu_x)}{a_{S,x}g_x}\approx 1.45\cdot10^{-1}.
    \end{aligned}
\end{equation*}
\begin{figure}[h]
    \centering
    
    \begin{subfigure}{0.49\textwidth}
    \centering
    \captionsetup{justification=centering,margin=0.5cm}   \includegraphics[width=1\textwidth]{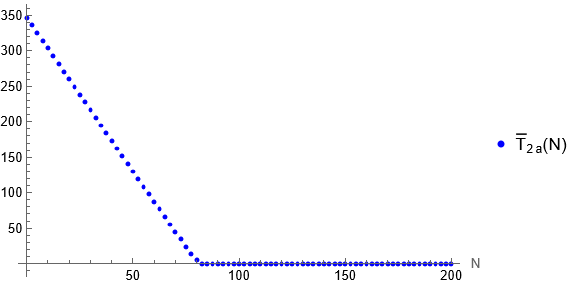}
    \caption{$\bar{T}_{2a}(N)=T(t^*;40,0,1000,N).$}
    \label{fig:8-a}
    \end{subfigure}
    \hfill
    \begin{subfigure}{0.49\textwidth}
    \centering
    \captionsetup{justification=centering,margin=0.5cm}
\includegraphics[width=1\textwidth]{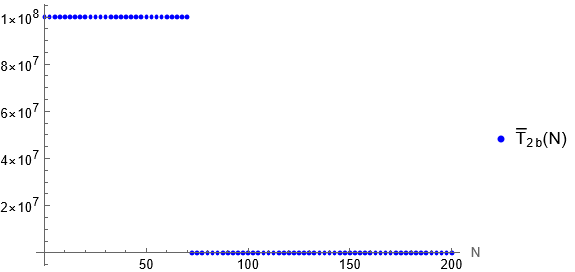}
    \caption{$\bar{T}_{2b}(N)=T(t^{**};40,0,1000,N).$}
    \label{fig:8-b}
    \end{subfigure}
    
    \begin{subfigure}{0.49\textwidth}
    \centering
    \captionsetup{justification=centering,margin=0.5cm}   \includegraphics[width=1\textwidth]{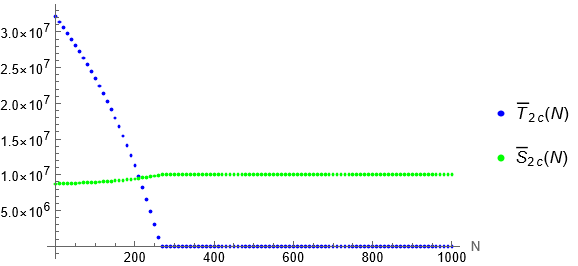}
    \caption{$\bar{T}_{2c}(N)=T(t^*;40,1,1000,N),$ $\bar{S}_{2c}(N):=S(t^*;40,1,1000,N).$}
    \label{fig:8-c}
    \end{subfigure}
    \hfill
    \begin{subfigure}{0.49\textwidth}
    \centering
    \captionsetup{justification=centering,margin=0.5cm}
\includegraphics[width=1\textwidth]{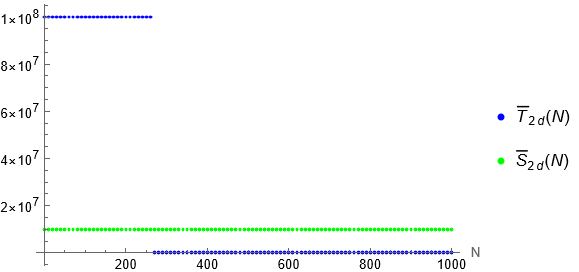}
    \caption{$\bar{T}_{2d}(N)=T(t^{**};40,1,1000,N),$ $\bar{S}_{2d}(N):=S(t^{**};40,1,1000,N).$}
    \label{fig:8-d}
    \end{subfigure}

    \begin{subfigure}{0.49\textwidth}
    \centering
    \captionsetup{justification=centering,margin=0.5cm}   \includegraphics[width=1\textwidth]{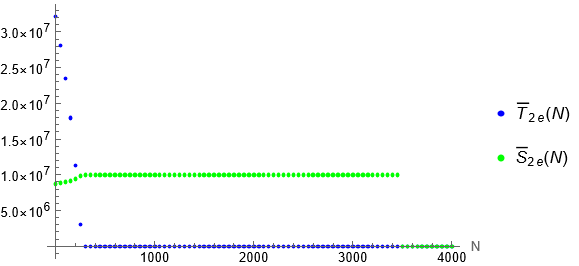}
    \caption{$\bar{T}_{2e}(N)=T(t^{*};40,1,1000,N),$ $\bar{S}_{2e}(N):=S(t^{*};40,1,1000,N).$}
    \label{fig:8-e}
    \end{subfigure}
    \hfill
    \begin{subfigure}{0.49\textwidth}
    \centering
    \captionsetup{justification=centering,margin=0.5cm}
\includegraphics[width=1\textwidth]{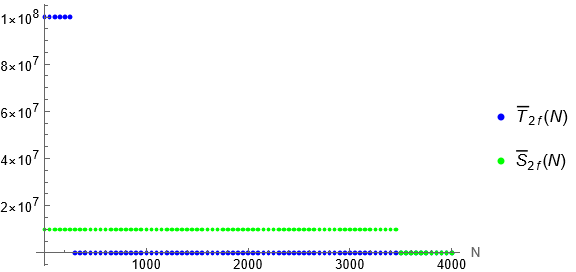}
    \caption{$\bar{T}_{2f}(N)=T(t^{**};40,1,1000,N),$ $\bar{S}_{2f}(N):=S(t^{**};40,1,1000,N).$}
    \label{fig:8-f}
    \end{subfigure}
    
    \caption{Plots of the values of solutions for small initial cancer cell populations after a given time depending on N.}
    \label{fig:8}
\end{figure}
The above inequalities are satisfied, in particular, for $N$ greater than approximately $1734.16.$ Of course, for some initial conditions, lower values of $N$ may be sufficient to bring the number of cancer cells to $0$. Figure~\ref{fig:8-a} shows that for $T_0=40, S_0=0, x_0=1000$ even $N\approx 80$ makes the number of cancer cells close to $0$ after $3$ months of treatment, while $N\approx 60$ is able to achieve the same goal after $3$ years of treatment; see Fig.~\ref{fig:8-b}. Note that if the treatment administered is not strong enough to make the cancer cell population converge to $0$ in $3$ months, then the number of cancer cells will still grow to reach maximum capacity, although slower. 

The situation changes dramatically when we introduce even a single cancer stem cell into the system, while leaving $T_0$ and $x_0$ unchanged. In fact, for $T_0=40,S_0=1,x_0=1000$ the value of $N$ greater than $220$ is necessary to make the number of ordinary cancer cells close to $0$ after 3 months of treatment, see Fig.~\ref{fig:8-c}. Figure \ref{fig:8-d} shows that the value of $N$ necessary to achieve the same goal after $3$ years of treatment remains essentially the same. On the other hand, much stronger immunotherapy, $N$ larger than approximately $3500,$ is required to make the number of cancer stem cells converge to $0,$ see Figs.~\ref{fig:8-e},~\ref{fig:8-f}.

\begin{figure}[h]
    \centering
    
    \begin{subfigure}{0.49\textwidth}
    \centering
    \captionsetup{justification=centering,margin=0.5cm}
\includegraphics[width=1\textwidth]{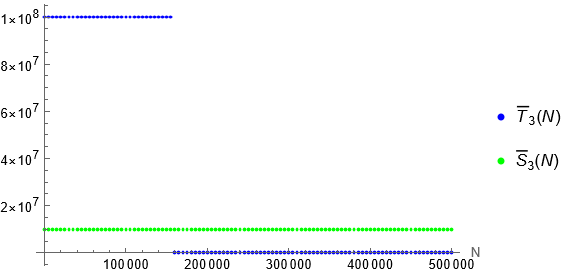}
    \caption{}
    \label{fig:3-a}
    \end{subfigure}
    \hfill
    \begin{subfigure}{0.49\textwidth}
    \centering
    \captionsetup{justification=centering,margin=0.5cm}
\includegraphics[width=1\textwidth]{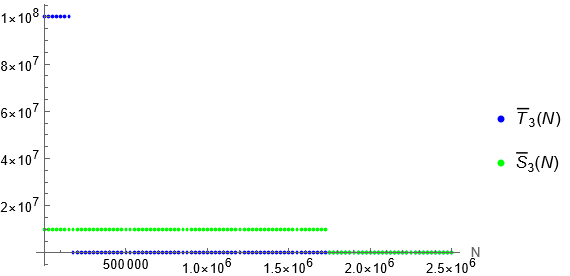}
    \caption{}
    \label{fig:3-b}
    \end{subfigure}
    
    \caption{Plots of the values of solutions for larger initial cancer cell populations after a given time depending on N -- $\bar{T}_{3}(N)=T(t^{**};2\cdot10^4,10^3,10^3,N),$ $\bar{S}_{3}(N)=S(t^{**};2\cdot10^4,10^3,10^3,N).$}
    \label{fig:3}
\end{figure}

Figure \ref{fig:3} shows that the situation is not qualitatively different for slightly larger initial conditions -- the value of $N$ that makes the number of CSCs converge to $0$ in $3$ years is more than ten times larger than the value of $N$ sufficient to achieve the same goal for ordinary cancer cells.

\begin{figure}[h]
    \centering
    
    \begin{subfigure}{0.49\textwidth}
    \centering
    \captionsetup{justification=centering,margin=0.5cm}
\includegraphics[width=1\textwidth]{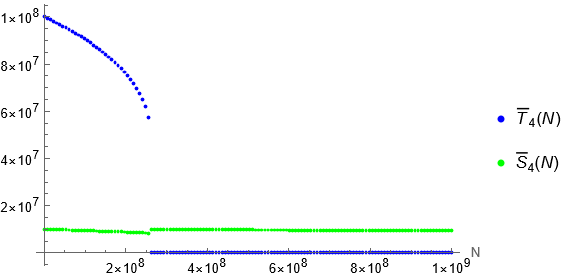}
    \caption{}
    \label{fig:4-a}
    \end{subfigure}
    \hfill
    \begin{subfigure}{0.49\textwidth}
    \centering
    \captionsetup{justification=centering,margin=0.5cm}
\includegraphics[width=1\textwidth]{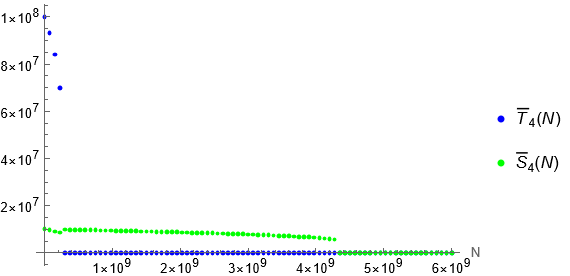}
    \caption{}
    \label{fig:4-b}
    \end{subfigure}
    
    \caption{Plots of the values of solutions for maximal initial cancer cell populations after a given time depending on N -- $\bar{T}_{4}(N)=T(t^{**};K_T,K_S,x^*(K_T,K_S),N),$ $\bar{S}_{4}(N)=S(t^{**};K_T,K_S,x^*(K_T,K_S),N).$}
    \label{fig:4}
\end{figure}

In Corollary \ref{wn:nie_ma_TSx} we state conditions on $N$ which guarantee that there is no strictly positive steady state exists, i.e.~if the condition from the Corollary \ref{wn:nie_ma_TSx} is met, then there is no such steady state where cancer cell populations of both types survive. For our specific parameter values these conditions read
\begin{equation*}
    \begin{gathered}
    G_T(N)\geqslant\frac{r_T(g_x+a_{x,T}K_T + a_{x,S}K_S+\mu_x)(K_T+1)}{a_{T,x}g_x+\mu_x}\approx 1.464\cdot 10^5,\\
    G_S(N)\geqslant\frac{r_S(g_x+a_{x,T}K_T + a_{x,S}K_S+\mu_x)(K_S+1)}{a_{S,x}g_x+\mu_x}\approx 1.464\cdot 10^6.
    \end{gathered}
\end{equation*}
Inequalities above hold, in particular, for $N$ greater than approximately $1.751\cdot10^{10}$. Figure \ref{fig:4} seems to suggest that in fact much smaller $N$ is sufficient to eradicate all cancer cells even with maximal initial conditions $T_0=K_T, S_0=K_S, x_0=x^*(K_T,K_S).$ Still, constant treatment of $N\approx4.2\cdot10^9,$ which as per Fig.~\ref{fig:4-b} appears to ensure full recovery after $3$ months, may not be implementable in reality.

\section{Conclusion}

In \cite{Abernathy} Abernathy and Burke extended the GBM immunotherapy model proposed in \cite{Kronik} and further analyzed in \cite{Kogan}, to account for the existence of CSCs. In this article, we applied a quasi-stationary approximation to this extended model, which yielded a simplified system with a reduced number of ODEs. We then introduced functions $G_T$ and $G_S$ that reflect the influence of immunotherapy on the dynamics of tumor cells and CSCs, respectively. We analyze the properties of these functions and show that they are strictly increasing, unbounded from above, and that $G_T$ is greater than $G_S$ for all positive values of $N$ (the total number of lymphocytes in the system at any given time), i.e.~that the treatment has a stronger impact on tumor cells than CSC. 

We then proceeded to analyze existence and local asymptotic stability of two types of steady states -- the cure state (when no cancer cells survive the treatment) and the coexistence steady state (when some populations of both types of cancer cells survive) -- depending on $N$. In particular, we found sufficient conditions for the local asymptotic stability of the cure state and for the non-existence of the coexistence steady state. Finally, we illustrated our findings with numerical simulations performed for particular values of parameters appearing in the model. 

A natural continuation of this work would be to relax the assumption that the therapy is constant in time and instead consider a treatment in the form of periodic impulses.
\\
\\


\begin{thebibliography}{30}

    \bibitem{GBM1}{
 {Q.T. Ostrom, H. Gittleman, P. Farah, A. Ondracek, Y. Chen, Y. Wolinsky, N.E. Stroup, C. Kruchko, J.S. Barnholtz-Sloan}.
 {{CBTRUS} Statistical Report: Primary Brain and Central Nervous System Tumors Diagnosed in the United States in 2006-2010},
 {\em Neuro-Oncology}, {2013},  {\bf 15}({suppl 2}),
 {ii1--ii56}.
}
    \bibitem{Hetero}{
 {A. Becker,  B. Sells, S. Haque, A. Chakravarti}.
 {Tumor Heterogeneity in Glioblastomas: From Light Microscopy to Molecular  Pathology},  {Cancers}, {2021},  {13}({4}), {761}.
    }
    \bibitem{Supress}{
{B.T. Himes, P.A. Geiger, K. Ayasoufi, A.G. Bhargav, D.A. Brown, I.F. Parney}.
 {Immunosuppression in Glioblastoma: Current Understanding and Therapeutic Implications}, {Frontiers in Oncology}, {2021},{\bf 11}.
    }
    \bibitem{GBM2}{
{Jigisha P. Thakkar and Therese A. Dolecek and Craig Horbinski and Quinn T. Ostrom and Donita D. Lightner and Jill S. Barnholtz-Sloan and John L. Villano}.
 {Epidemiologic and Molecular Prognostic Review of Glioblastoma}, {\em Cancer Epidemiology,  Biomarkers \& Prevention},  {2014}, {\bf 23}({10}),
 {1985--1996}.
    }
    \bibitem{GBM3}{
 {C. Fernandes, A. Costa, L. Os\'orio, R.C. Lago, P. Linhares, B. Carvalho, C. Caeiro}. 
 {Current Standards of Care in Glioblastoma Therapy},
in {\bf Glioblastoma}, {197--241},
 {Codon Publications},  {2017}.
    }
    \bibitem{Immuno}{
 {N. Desbaillets, A.F. Hottinger}.  {Immunotherapy in Glioblastoma: A Clinical Perspective}, {\em Cancers},  {2021}, {\bf 13}({15}), {3721}.
    }
    \bibitem{Kronik}{
{N. Kronik}, {Y. Kogan}, {V. Vainstein}, {Z. Agur}.
{Improving alloreactive {CTL} immunotherapy for malignant gliomas using a simulation model of their interactive dynamics}, {\em Cancer Immunology,  Immunotherapy},  {2007}, {\bf 57}({3}), {425--439}.
    }
    \bibitem{Kogan}{
{Y. Kogan, U. Fory\'s, O. Shukron, N. Kronik, Z. Agur}.
{Cellular immunotherapy for high grade gliomas: mathematical analysis deriving efficacious infusion rates based on patient requirements}, {\em SIAM Journal on Applied Mathematics}, 2010, {\bf 70}(6), 1953--1976.
    }
    \bibitem{Abernathy}{
{K. Abernathy}, {J. Burke}. 
{Modeling the Treatment of Glioblastoma Multiforme and Cancer Stem Cells with Ordinary Differential Equations}, {\em Computational and Mathematical Methods in Medicine},  {2016}, ID 1239861.
    }
    \bibitem{Stem}{
 {B.T. Tan}, {C.Y. Park}, {L.E. Ailles}, {I.L. Weissman}.
 {The cancer stem cell hypothesis: a work in progress},
 {\em Laboratory Investigation}, {2006}, {\bf 86}({12}), {1203--1207}.
    }

\end{thebibliography}
\end{document}